\documentclass[twocolumn]{autart}    

\usepackage{graphicx}          
\usepackage{comment}
\usepackage{dsfont}
\usepackage{bm}
\usepackage{mathrsfs}
\usepackage{amsfonts}
\usepackage{amssymb}
\usepackage{amsmath}
\usepackage{subfigure}
\usepackage{enumerate}
\usepackage{cite}
\usepackage{xcolor}
\usepackage{algpseudocode}
\usepackage{algorithmicx,algorithm}
\usepackage{diagbox}

\usepackage[english]{babel}
\definecolor{mblue}{rgb}{0,0.267,0.486}

\begin{document}
\begin{frontmatter}

\title{A Structure-Tensor Approach to Integer Matrix Completion \\ with Applications to Differentiated Energy Services\thanksref{footnoteinfo}}

\thanks[footnoteinfo]{This work was partially supported by the Research Grants Council of Hong Kong Special Administrative Region, China, under the Theme-Based Research Scheme T23-701/14-N.}

\author[HKUST]{Yanfang Mo}\ead{ymoaa@connect.ust.hk},    
\author[HKUST]{Wei Chen}\ead{wchenust@gmail.com},               
\author[HKU]{Sei Zhen Khong}\ead{szkhong@hku.hk},
\author[HKUST]{Li Qiu}\ead{eeqiu@ust.hk}  

\address[HKUST]{Department of Electronic and Computer Engineering, Hong Kong University of Science and Technology, \\Clear Water Bay, Kowloon, Hong Kong, China}  
\address[HKU]{Department of Electrical and Electronic Engineering, University of Hong Kong, Pokfulam, Hong Kong, China}  


\begin{abstract}                          
Efficient resource allocation is one of the main driving forces of human civilizations. Of the many existing approaches to resource allocation, matrix completion is one that is frequently applied. In this paper, we investigate a special type of matrix completion problem concerning the class of $(0,1)$-matrices with given row/column sums and certain zeros prespecified. We provide a necessary and sufficient condition under which such a class is nonempty. The condition is stated in the form of the nonnegativity of a structure tensor constructed from the information regarding the given row/column sums and fixed zeros. Moreover, we show that a more general matrix completion problem can be studied in a similar manner, namely that involving the class of nonnegative integer matrices with prescribed row/column sums, predetermined zeros, and different bounds across the rows. To illustrate the utility of our results, we apply them to demand response applications in smart grids. Specifically, we address two adequacy problems in differentiated energy services, namely, the problems of supply/demand matching and minimum purchase profile. 


\end{abstract}

\end{frontmatter}

\section{Introduction}
Resource allocation is well documented in history and ubiquitous in real life. Human ancestors distribute food to survive and thrive. The victorious generals share captured valuables with soldiers to enhance troop morale. Factories manage manpower and materials for profit maximization. Students apportion their time and energy to various courses and activities. We refer to~\cite{boggs1992resource,van1986acquisition} for more detailed accounts of the significance of resource allocation in our lives. 

The problem of resource allocation is well studied in the literature. Many ways to model and solve the problem can be found. For example, network flow theory associates a proper allocation with a flow in a network~\cite{ahuja1993network}. Resources are distributed in auction theory by mechanism design~\cite{krishna2009auction}. Transportation theory is concerned with transporting a distribution of resources to another one while minimizing a certain transportation cost~\cite{villani2003topics}. Operations research approaches the problem of resource allocation with tools ranging from portfolio selection, computer scheduling, production planning, and apportionment~\cite{ibaraki1988resource}. A very special class of resource allocation problems has been studied in the area of leader selection and token allocation~\cite{giua2002firing,richert2013optimal}. For more pioneering works on resource allocation, see~\cite{chasparis2016design,chen2017optimal,ghosh1958input,jain2010efficient,martin2005time}. 

Matrix completion is another widely used tool for resource allocation. A matrix presents a useful mathematical depiction of allocating resources from their sources to destinations. In particular, the row indices of the matrix may represent the scattered locations of sources while each row sum corresponds to the total amount of resources at that particular location. Likewise, the column indices and sums may represent the same for the destinations. The viability of the resource allocation problem is therefore equivalent to the existence of a matrix satisfying the given row and column sums. Such a matrix is often referred to as a transportation matrix, since each matrix element indicates the quantity of resources transported from a certain source to a certain destination. This can be viewed as a discrete version of the classic optimal mass transport problem. Apart from attributing the locations of resources to the matrix indices, many other properties such as time intervals, varieties, purposes, and hierarchical structures may be used.


In this paper, we consider constrained nonnegative integer matrix completion problems. These problems, which are NP-hard in general, are significantly more difficult than completing a matrix with real elements, which may be efficiently solved by using linear programming~(LP) directly. Our formulation is motivated by the commonplace presence of resources that are indivisible, such as a human, a delivery package, and a biochemical process that cannot be resumed once interrupted. In these examples, the elements in the matrix to be completed need to be taken as nonnegative integers.

As a starting point, we investigate the class of $(0,1)$-matrices with given row/column sums and certain zeros prespecified and provide a necessary and sufficient condition under which such a class is nonempty. Variants of this problem have been widely studied over the past century. In the absence of prespecified elements, the seminal Gale-Ryser theorem characterizes the problem via a majorization condition involving only the row/column sums~\cite{gale1957theorem,ryser1957combinatorial}. Anstee~\cite{anstee1982properties} considers the case where there is at most one prespecified zero in each column. The existence condition of a triangular $(0,1)$-matrix is elaborated in~\cite{anstee1982triangular}. The case where all the fixed zeros form a block at a corner is examined in~\cite{brualdi2003matrices}. One of our previous papers deals with the case where the positions of the fixed zeros constitute a staircase pattern~\cite{chen2016constrained}. Comprehensive surveys along the above line of research are available in~\cite{brualdi2006combinatorial,olkin2016inequalities}. In addition, if the matrix is square, then such a $(0,1)$-matrix completion problem is in essence the graph realization problem in graph theory and related works can be found in~\cite{berger2011dag,chen1966realization,erdosgallai1960degree}.

The aforementioned works lay the foundations for the results in this paper. Specifically, we define a structure tensor to characterize the non-emptiness of the class of~$(0,1)$-matrices with given row/column sums and predetermined zeros. In the case where the~$(0,1)$-matrix class is nonempty, we show that we can find an element of the class via the structure-tensor approach efficiently when the pattern of fixed zeros presents a desired structure. As a generalization, we also examine via the structure-tensor approach the more general matrix completion problem involving nonnegative integer matrices with prescribed row/column sums, predetermined zeros, and upper bounds across the rows.

As an illustration of the results developed in this paper, we apply them to the analysis and design of differentiated energy services. Such services have been put forward as a form of demand response, whose purpose is to exploit the flexibilities in demands to alleviate the burdens on supplies in smart grids~\cite{siano2014demand}. The electricity services herein are neither homogeneous products sold at a unit price nor plug-and-play charging processes. They are differentiated by different flexibility levels of the unique charging properties of smart loads. For example, the charging of an electric vehicle can be deferrable, intermittent, and modulated. That is, the charging may not start immediately and does not have to be continuous as long as it is completed within a specified time frame. Thus, we can coordinate the charging processes of these flexible loads to better maintain the efficiency of the power systems. In this paper, we concentrate on two adequate problems in differentiated energy services.


The first problem is about the supply/demand matching~--~under what situations can the supply fulfill all the load requirements? It will be shown that finding such an adequacy condition is equivalent to characterizing the existence of a constrained $(0,1)$-matrix, called the feasible power allocation matrix, which indicates how the given supply is allocated to conform to the demand requirements. Even though the existence can be verified by solving a special class of integer linear programs~(ILPs) with existing algorithms, the structure-tensor approach in this paper is more informative. It offers the physical interpretation that the supply tails should always dominate the demand tails whenever the supply is adequate.

Next, when the supply is inadequate, a follow-up question asks what the minimum supplementary purchase is. A simple algorithm is developed to achieve an optimal purchase profile of the minimum amount (i.e., the adequacy gap), with the aid of the structure tensor. By further making use of the Gale-Ryser theorem, a refined algorithm is proposed to solve the minimum purchase profile problem more efficiently.

Finally, we take a step forward towards the more complicated rate-constrained differentiated energy services, where a load may be charged at an integer multiple of the base rate at each time slot, between zero and a certain ceiling charging rate. In the previous setup, every load can be charged at the base rate at most. However, in this case, the ceiling charging rate is load-dependent but invariant to the load's service time. Two adequacy problems described previously are revisited in this new framework. Mathematically, they lead to more general matrix completion problems, which involve the class of nonnegative integer matrices with prescribed row/column sums, predetermined zeros, and different bounds across the rows. Similarly to before, we propose a generalized structure tensor to address the problems of supply/demand matching and minimum purchase profile.

Here is the outline of the paper. Preliminary knowledge is presented in Section~$2$. We study the class of $(0,1)$-matrices with given row/column sums and fixed zeros in Section~$3$. The allowable patterns of fixed zeros generalize those studied in~\cite{chen2016constrained}, which are limited to taking the shape of a staircase. In Section~$4$, we consider the adequacy problems in differentiated energy services, namely, the supply/demand matching and the minimum purchase profile problems. We assume in this section that each load can either be charged at the base charging rate or receive no power at each time slot. Partial results have been recorded in the conference paper~\cite{mo2017differentiated}. In Section~$5$, we investigate the more complicated rate-constrained differentiated energy services and nonnegative integer matrix completion problems. Finally, we conclude this paper and propose future work in Section~$6$.

\textsl{Notation}\\
Let $\mathbb{R}$, $\mathbb{R}^+$, and $\mathbb{N}$ denote the set of real numbers, positive real numbers, and nonnegative integers respectively. Other sets are denoted by capital calligraphic letters. Considering a subset~$\mathcal{X}$ of a set~$\mathcal{A}$, we use $\mathcal{A}\backslash\mathcal{X}$ to denote the set~$\mathcal{Y}$ such that $\mathcal{X}\cup \mathcal{Y}=\mathcal{A}$ and $\mathcal{X}\cap \mathcal{Y}=\emptyset$. Consequently, the pair of sets~$(\mathcal{X},\mathcal{A}\backslash\mathcal{X})$ is a partition of the set~$\mathcal{A}$. Tensors, including matrices, are denoted by italic capital letters, except for~$N$ and~$T$. We reserve~$O$ or~$E$ to represent matrices with all the elements being zeros and ones respectively, whose dimensions are inferred from the context. The $n$th row or $j$th column of a matrix~$A$ is respectively specified by $A(n,:)'$ or $A(:,j)$. For two matrices of the same size, $A$ and $B$, we write~$A\leq B$ if $A(n,j)\leq B(n,j)$ for every~$n$ and $j$. Let $\mathds{1}(A)$ map a matrix~$A$ to a $(0,1)$-matrix of the same size by changing every nonzero element of~$A$ to one and $\mathds{1}(\mathfrak{A})$ map a true~(\emph{resp.} false) assertion~$\mathfrak{A}$ to one~(\emph{resp.} zero). The function~$\|\cdot\|_1$ denotes the H\"{o}lder 1-norm of a real matrix or vector, which is the summation of absolute values of all the elements. Let~$[a]^+$ denote the maximum of zero and a real number~$a$. 

\section{Preliminary}
\subsection{Majorization}\label{apdpre}

Majorization plays an important role in the theory of inequalities. 
Following are its basic concepts and more details can be found in the monograph \cite{olkin2016inequalities}.

For a vector $\bm{x}~=~[x_1~x_2~\cdots~x_N]'$, we denote its non-increasing rearrangement as $x_{[1]}\geq x_{[2]}\geq \dots \geq x_{[N]}$.\
\begin{defn}
  For $\bm x, \bm y \in \mathbb{R}^N$, we write $\bm x\prec^w \bm y$ if $$\sum\nolimits_{j=n}^{N}x_{[j]}\geq \sum\nolimits_{j=n}^{N}y_{[j]},\ \forall n=1, 2,\dots,N;$$\\
  and write $\bm x\prec_w \bm y$ if $$\sum\nolimits_{j=1}^{n}x_{[j]}\leq \sum\nolimits_{j=1}^{n}y_{[j]},\ \forall n=1, 2,\dots,N.$$
  In the former, $\bm x$ is said to be weakly supermajorized by~$\bm y$, while in the latter, $\bm x$  is said to be weakly submajorized by~$\bm y$. We write $\bm x \prec \bm y $ and say that $\bm x$ is majorized by $\bm y$, if we further have $\sum_{n=1}^{N}x_n\!=\!\sum_{n=1}^{N}y_n$.
\end{defn}

Clearly, if $\bm x\prec \bm y$ and $\bm y \prec \bm x$, then the non-increasing arrangements of the two vectors are the same. Thus, by restricting ourselves to vectors in~$\mathbb{R}^N$ with non-increasing elements, we can regard majorization as a partial order~(satisfying the reflexivity, antisymmetry, and transitivity properties).

If a vector consists of nonnegative integers only, then it can be treated as a partition of a certain integer, which is the total sum of all the elements. Considering a partition~$\bm x \in \mathbb{N}^N$, we define its partition conjugate, denoted by~$\bm x^*$, by setting $x_j^*$ as the number of elements of $\bm x$ that are no less than $j$, i.e.,
\begin{alignat*}{8}
&x_j^*=\sum\nolimits_{n=1}^{N}\mathds{1}(x_n\geq j),\ \forall j\in \mathbb{N}/\{0\}.
\end{alignat*}
From the definition, $\bm x^*$ is organized in a non-increasing order and $x_j^*=0$ when $j$ is greater than the largest element of~$\bm x$. Note that it is a general practice that we can adjust the number of zeros such that $\bm x^*$ possesses the size inferred from the context. In addition, the partition conjugate has the following property:
$$x^{**}_n=x_n, n = 1,2, \dots,N.$$
Furthermore, we emphasize that the partition conjugate acts as a bridge between the two concepts of ``weak'' majorization, since $\bm x \prec_w \bm y \Leftrightarrow \bm y^* \prec^w \bm x^*.$

\subsection{Network Flow}
We herein give a brief introduction to network flow theory. A thorough review can be found in~\cite{korte2012combinatorial} and~\cite{trevisan2011combinatorial}.

A directed graph~$G$ consists of a vertex set~$\mathcal{V}$ and a set~$\mathcal{E}$ of arcs~(directed edges), written as $G=(\mathcal{V},\mathcal{E})$. If there exists an arc oriented from a vertex~$u$ to a vertex~$v$, then we denote it by~$(u,v)$ and call $u$~(\emph{resp.}~$v$) the head~(\emph{resp.} tail) of the arc. The indegree of a vertex is the number of arcs directed into the vertex, while the outdegree of a vertex is the number of arcs directed out of the vertex. A vertex is called a source node if its indegree is zero, while a vertex is called a sink node if its outdegree is zero. An $s$-$t$ network~$(G,s,t,c)$ refers to a directed graph~$G=(\mathcal{V}, \mathcal{E})$ containing two distinguished nodes: $s\in \mathcal{V}$~(a source node) and $t\in \mathcal{V}$~(a sink node), together with a capacity function~$c$: $\mathcal{E}\rightarrow \mathbb{R}^+$. The capacity function~$c$ maps each arc~$(u,v)\in \mathcal{E}$ to a positive number~$c(u,v)$, which is called the capacity of the arc. In addition, vertices of~$G$ other than $s$ and $t$ are called the internal nodes of the $s$-$t$ network~$(G,s,t,c)$. 

The concept of a flow is significant in the study of $s$-$t$ networks. Its definition is given in the following.
\begin{defn}
    A flow $f$ in an $s$-$t$ network $(G,s,t,c)$ is a nonnegative real-valued function defined on the arc set: $\mathcal{E}\rightarrow \mathbb{R}^+$, subject to the following two constraints:
  \begin{enumerate}
    \item Capacity constraint: for every arc~$(u,v) \in \mathcal{E}$, $f(u,v)\leq c(u,v);$
    \item Conservative law: for every internal node~$u \in \mathcal{V}$, $\sum_{(v,u)\in \mathcal{E}}f(v,u)=\!\!\sum_{(u,v)\in \mathcal{E}}f(u,v).$
  \end{enumerate}
\end{defn}
 A flow is said to be integral if the function values are all integers. Moreover, we define the value of a flow~$f$ by $$|f|=\sum_{v\in V, (s,v)\in \mathcal{E}}f(s,v)=\sum_{u\in V, (u,t)\in \mathcal{E}}f(u,t).$$ In order to find an $s$-$t$ flow of the maximum value, we recall a classic problem in network flow theory, i.e., the maximum flow problem. The following theorem characterizes the existence of an integral flow which solves a special class of maximum flow problems\cite{dantzig1955max}.
\begin{thm}[Integral Flow]
  If all the capacities of an $s$-$t$ network are integers, then there exists an integral flow which has the maximum value of a flow.
\end{thm}
A dual concept of an $s$-$t$ flow is the $s$-$t$ cut defined below.
\begin{defn}
A cut in an $s$-$t$ network $(G,s,t,c)$ is a partition~$(\mathcal{X},\mathcal{V}/\mathcal{X})$ of $\mathcal{V}$ such that $s\in \mathcal{X}$ and $t\in \mathcal{V}/\mathcal{X}$.
\end{defn}
The capacity of the cut~$(\mathcal{X},\mathcal{V}/\mathcal{X})$ is given by $$c(\mathcal{X},\mathcal{V}/\mathcal{X})=\sum\nolimits_{u\in\mathcal{X}, v\in \mathcal{V}/\mathcal{X}, (u,v)\in \mathcal{E}}c(u,v).$$
Note that the dual of the maximum flow problem is to find an $s$-$t$ cut of the minimum capacity. This leads to a central theorem of network flow theory~\cite{ford1956maximal} as below.
\begin{thm}[Max-Flow-Min-Cut]
  In an $s$-$t$ flow network, the maximum value of a flow equals the minimum capacity of a cut.
\end{thm}
In fact, the Max-Flow-Min-Cut theorem is an application of the duality theorem of linear inequality theory. To explain further, the above theorem can be restated in terms of the following three equivalent statements
\begin{enumerate}
  \item There is a cut~$(\mathcal{X},\mathcal{V}/\mathcal{X})$ whose capacity is equal to the value of a flow~$f$ in the $s$-$t$ network.
  \item The flow~$f$ possesses the maximum value.
  \item The cut~$(\mathcal{X},\mathcal{V}/\mathcal{X})$ possesses the minimum capacity.
\end{enumerate}

\section{$(0,1)$-Matrix Completion}
In this section, we consider the class of $(0,1)$-matrices with given row/column sums and fixed zeros. A necessary and sufficient condition is provided under which the matrix class is nonempty. The condition is stated in the form of the nonnegativity of a structure tensor. Furthermore, when the matrix class is not empty and the pattern of fixed zeros presents a desired structure, we demonstrate the use of the tensor condition to find a matrix in the matrix class efficiently.
\subsection{Mathematical Model}
Given an~$N\times T$ $(0,1)$-matrix, denote the column and row sums respectively by two nonnegative integer vectors:
$$\bm h = [h_1~h_2~\cdots~h_T]'\text{ and }\bm r = [r_1~r_2~\cdots~r_N]'. $$
They are respectively called the \emph{column sum vector} and \emph{row sum vector}. The prescribed zeros are specified by the pattern matrix~$F$, which is a $(0,1)$-matrix of size~$N\times T$. Denote by~$\mathcal{A}(\bm h,\bm r, F)$ the class of $N\times T$ matrices with the column sum vector~$\bm h$, the row sum vector~$\bm r$, and the pattern matrix~$F$. Specifically, a matrix~$A\in \mathbb{N}^{N\times T}$ belongs to the matrix class~$\mathcal{A}(\bm h,\bm r, F)$ if and only if
\begin{eqnarray}
  A(n,j)\!\in\! \{0,1\},\ \forall n=1,2,\dots,N~\&~j=1,2,\dots,T;\label{c0}\\
  \|A(n,:)'~\|_1=r_n,\ \forall n=1,2,\dots,N; \label{c1} \\
 \|A(:,j)\|_1= h_j,\ \ \forall j=1,2,\dots,T;\label{c2} \\
   O \leq A \leq F.\label{c3}
\end{eqnarray}
Our main objective is to characterize the conditions under which the matrix class~$\mathcal{A}(\bm h,\bm r, F)$ is not empty. A necessary condition for the existence of such a matrix~$A$ is given by~$\|\bm h\|_1=\|\bm r\|_1$, but it is not sufficient.

From an optimization perspective, $\mathcal{A}(\bm h,\bm r, F)$ is the feasible region described by constraints~(\ref{c0})--(\ref{c3}). Whether there exists a matrix in~$\mathcal{A}(\bm h,\bm r, F)$ is equivalent to the feasibility of the problem constrained by~(\ref{c0})--(\ref{c3}). This leads to an ILP. Nevertheless, the feasible region described by constraints~(\ref{c1})--(\ref{c3}) is actually an integer polyhedron given by the convex hull of~$\mathcal{A}(\bm h,\bm r, F)$. Hence, the feasibility problem constrained by~(\ref{c1})--(\ref{c3}) admits a solution which is also feasible for that constrained by~(\ref{c0})--(\ref{c3}). As a result, we can check the non-emptiness of~$\mathcal{A}(\bm h,\bm r, F)$ by solving an associated LP. In addition, we will show later that this problem has an equivalent network-flow formulation and thus we can check whether~$\mathcal{A}(\bm h,\bm r, F)$ is empty in polynomial time via network-flow algorithms.

In addition to these aforementioned numerical methods, an alternative approach involves deriving a collection of inequalities from the structural information~$(\bm h,\bm r, F)$ so as to verify the non-emptiness of~$\mathcal{A}(\bm h,\bm r, F)$. This was independently initiated by D.~Gale and H.~J.~Ryser to deal with~$\mathcal{A}(\bm h,\bm r, E)$, where there is no constraint on the zero-pattern~\cite{gale1957theorem,ryser1957combinatorial}. Thereafter, a number of existence conditions have been derived when certain zero-patterns are present, such as~$F$ being a triangular matrix~\cite{anstee1982triangular}. These conditions also help design specialized algorithms to find matrices in the described class~$\mathcal{A}(\bm h,\bm r, F)$\cite{brualdi2006combinatorial}. By continuing this line of research, we herein generalize the classic Gale-Ryser theorem for the case with a general~$F$ via a structure-tensor approach.

Before moving to the key result, we introduce more concepts regarding the pattern matrix which are necessary for the description of our structure-tensor condition. Considering an arbitrary pattern matrix~$F$, we say $F$ can be described by $\lambda+1$ special column indices if we can find special column indices, denoted in the natural order by $T_0<T_1<T_2<\cdots<T_{\lambda-1}<T_\lambda,$ such that
\begin{equation}\label{despattern}
  F(:,T_i+1)=F(:,T_i+2)=\cdots=F(:,T_{i+1}),
\end{equation}
for $i=0,1,\dots,\lambda-1.$ Clearly, $T_0=0$, $T_\lambda =T$ and $\lambda\geq 1$. Although our structure-tensor technique can be applied to every kind of pattern matrices, we emphasize that it is particularly useful when the pattern matrix~$F$ has a rather large number of rows and can be described by a relatively small~$\lambda$. For notational convenience, we assume the following monotonicity throughout this paper:
 \begin{equation}\label{monoto}
  h_{T_i+1}\geq h_{T_i+2}\geq\cdots\geq h_{T_{i+1}},\ \forall i=0,1,\dots,\lambda-1.
\end{equation}
\begin{rem}
  It is worth noting that our results in this paper are also suitable for the class of $(0,1)$-matrices with given row/column sums and predetermined zeros/ones. The reason is as follows. If there are fixed ones, we replace them by zeros and decrease the given row and column sums correspondingly. In doing so, we obtain a new class of $(0,1)$-matrices with predetermined zeros only. It is noticeable that there exists a one-to-one correspondence between the matrices in the original class and those in the new class.
\end{rem}

\subsection{A Necessary and Sufficient Condition}
\begin{figure}[t]
  \begin{minipage}[t]{0.5\linewidth}
    \centering
   \includegraphics[scale=0.7]{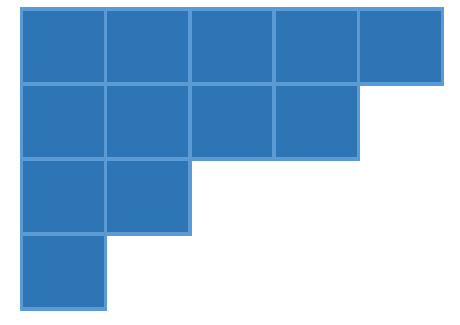}
  \end{minipage}%
  \vline
  \begin{minipage}[t]{0.5\linewidth}
    \centering
    \includegraphics[scale=0.7]{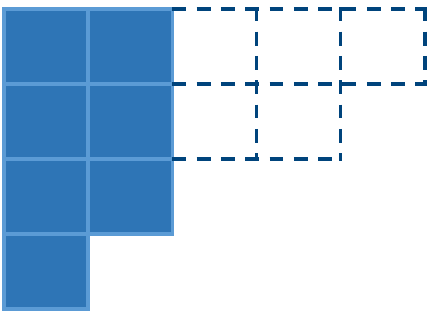}
  \end{minipage}
  \caption{A Young diagram of shape~$[5~4~2~1]'$.}
  \label{fig: YoungDiagram}
\end{figure}

Our structure-tensor condition is in essence a generalization of the Gale-Ryser theorem and we firstly reproduce this pioneering result in the following.
\begin{thm}[Gale-Ryser]\label{GaleRyser}
The matrix class~$\mathcal{A}(\bm h,\bm r, E)$ is nonempty if and only if $\bm h \prec \bm r^*$.
\end{thm}
Let us explain its algebraic details from a graphical perspective, by virtue of the Young diagram defined below. By convention, the shape of a Young diagram is denoted by~$(r_1,r_2,\ldots,r_N)$, where $r_1\geq r_2 \geq \cdots \geq r_N$. Without ambiguity, we write it as a vector in its non-increasing rearrangement to be consistent with other concepts like partition conjugate.
\begin{defn}
The Young diagram of shape~$[r_1~r_2~\cdots~r_N]'$ is a collection of left-justified cells aligned such that there are $r_i$ cells in the $i$th row, for $i=1,2,\dots,N$.
\end{defn}
An illustrative instance is given on the left of~Fig.~\ref{fig: YoungDiagram}. Considering the $j$th column of a Young diagram, there will be a cell at the $i$th row if and only if the $i$th element of the shape is no less than~$j$. Thus, we conclude that the number of cells in each column makes up the partition conjugate of the shape of a Young diagram. In this spirit, by dint of the Young diagram whose shape is the row sum vector~$\bm r$, the inequalities~$$\sum_{j>k}^{T}h_j \geq  \sum_{j>k}^{T}r^*_j, k=0,1,\dots,N,$$ suggest that the sum of the least~$N-k$ elements of the column sum vector~$\bm h$ should be no less than the number of cells in the corresponding~$N-k$ columns of the Young diagram. As illustrated by the right of~Fig.~\ref{fig: YoungDiagram}, if $T=5,k=2,$ and $r=[5~4~2~1]'$, then the least three elements of~$\bm h$ should be no less than five, which is the number of dashed cells. Thus, by writing~$\bm h \prec \bm r^*$, we actually count the dashed cells by column. In contrast, if we count the number of dashed cells by row, we attain another expression for the majorization inequality in the Gale-Ryser theorem, which combines $\|\bm h\|_1=\|\bm r\|_1$ with
\begin{equation}\label{OneOrderTensor}
   \sum_{j>k}^{T}h_j-\sum_{n=1}^{N}[r_n-k]^+\geq 0, k=1,2,\dots,T.
\end{equation}
At this point, we can define a structure tensor by imitating the left part of the expression~(\ref{OneOrderTensor}). In the Gale-Ryser case where $F=E$ and thus $\lambda=1$, we have only one index~(denoted by~$k$) and one corresponding summation from the column side. If the pattern matrix~$F$ can be described by~$\lambda+1$ special column indices~($\lambda\geq 1$), then we define a $\lambda$th-order tensor as 
\begin{equation*}\label{tensor0}
\begin{split}
  W_{k_1k_2\cdots k_{\lambda}}(\bm{h},\bm{r},F)\!= \! \! \sum_{j>k_1}^{T_1}\!h_j+\!\!\!\!\sum_{j>T_1+k_2}^{T_2}\!\!\!\!h_j+\cdots+\!\!\!\!\!\!\!\sum_{j>T_{\lambda-1}+k_{\lambda}}^{T_{\lambda}}\!\!\!\!h_j\\
  -\sum_{n=1}^{N}\left[r_n-\sum_{i=1}^{\lambda}k_iF(n,T_i)\right]^+, \end{split}
\end{equation*}
where~$k_i=0,1,\dots,T_i-T_{i-1}$, for $i=1,2,\dots,\lambda$. We call this tensor a structure tensor, since it is totally determined by the structural information~$(\bm h,\bm r,F)$. The size of the tensor is~$$\left(T_1-T_0+1\right)\times\left(T_2-T_1+1\right)\times\cdots\times\left(T_\lambda-T_{\lambda-1}+1\right),$$ which is solely determined by the pattern matrix~$F$. We write $W(\bm h, \bm r, F)\geq 0$, if every element of $W(\bm h, \bm r, F)$ is nonnegative.

\begin{exmp} \label{exmp1}
A pattern matrix~$F$ is presented on the left of Fig.~\ref{fig: concretestruc}, which signifies that $\lambda=2$, $T_0=0$, $T_1=1$, and $T_2=3$. Thus, given any row and column sums, the associated structure tensor should have the size $2\times 3$, where~$k_1=0,1$ and $k_2=0,1,2$. Specifically, considering $$\bm{\hat{h}}=[2~2~1]', \bm{\tilde{h}}=[1~2~2]'\text{, and }\bm r=[3~1~1]',$$ the values of the two tensors~$W(\bm{\hat{h}},\bm r, F)$ and~$W(\bm{\tilde{h}},\bm r, F)$ are presented on the right of Fig.~\ref{fig: concretestruc} in the order they are mentioned. In particular, we calculate~$W_{02}(\bm{\tilde{h}},\bm r, F)$ by
\begin{figure}[t]
\centering
\begin{minipage}[t]{0.15\linewidth}
    \centering
   \mbox{$\begin{bmatrix}
       1 & 1 & 1  \\
       1 & 1 & 1  \\
       1 & 0 & 0
     \end{bmatrix}$}
  \end{minipage} \vline
  \begin{minipage}[h]{0.83\linewidth}
    \centering
    \mbox{\begin{tabular}{|c|r@{\hspace{0.68em}}|r@{\hspace{0.68em}}|r@{\hspace{0.68em}}|}\hline
\diagbox[height=3em,width=3em]{$k_1~~$}{$~~~~~k_2$}& $~0$ & $~1$ & $~2$\\ \hline
$~0$ & $~0$ & $~0$ & $~0$\\ \hline
$~1$ & $~1$ & $~0$ & $~0$\\ \hline
\end{tabular}~\mbox{\begin{tabular}{|c|r@{\hspace{0.68em}}|r@{\hspace{0.68em}}|r@{\hspace{0.68em}}|}\hline
\diagbox[height=3em,width=3em]{$k_1~~$}{$~~~~~k_2$}& $~0$ & $~1$ & $~2$\\ \hline
$~0$ & $~0$ & $~0$ & \textcolor{mblue}{-$1$}\\ \hline
$~1$ & $~2$ & $~1$ & $~0$\\ \hline
\end{tabular}}}
  \end{minipage}
  \caption{A pattern matrix and two structure tensors.}
  \label{fig: concretestruc}
\end{figure}
\begin{align*}
  &\tilde{h}_1-[r_1-k_1-k_2]^+-[r_2-k_1-k_2]^+-[r_3-k_1]^+\\
 =~ &1-[3-0-2]^+-[1-0-2]^+-[1-0]^+=-1.
\end{align*}
Thus, we have~$W(\bm{\hat{h}},\bm r, F)\geq 0$ and $W(\bm{\tilde{h}},\bm r, F)\ngeq 0$.
\end{exmp}

As a generalization of the Gale-Ryser theorem, the following theorem makes use of the associated structure tensor and gives a necessary and sufficient condition for the non-emptiness of the $(0,1)$-matrix class with given row and column sums, together with a pattern matrix~$F$ specified by~$\lambda+1$ special column indices.
\begin{thm}\label{tensorone}
  The matrix class~$\mathcal{A}(\bm h, \bm r, F)$ is nonempty if and only if $W(\bm h,\bm r, F)\geq 0$ and $\|\bm h\|_1=\|\bm r\|_1$. 
\end{thm}
For Example~\ref{exmp1}, we can conclude by Theorem~\ref{tensorone} that the matrix class~$\mathcal{A}(\bm{\hat{h}},\bm r, F)$ is nonempty while~$\mathcal{A}(\bm{\tilde{h}},\bm r, F)$ is empty, which can be verified easily. Before giving a rigorous proof of Theorem~\ref{tensorone}, we present a useful lemma. 
\begin{lem}\label{keylem}
  Given two vectors $\bm{a}, \bm{b}\in \mathbb{R}^N$, it holds that:\\
  \centering $\max\limits_{p,\{l_1,l_2,\ldots,l_p\}}\sum\limits_{i=1}^{p}\left(a_{l_i}-b_{l_i}\right)=\sum\limits_{n=1}^{N}\left[a_n-b_n\right]^+.$
\end{lem}
\begin{pf}
  We divide the set~$\{a_{n}-b_{n}~|~n=1,2,\ldots,N\}$ into two separate subsets so that the term~$a_{n}-b_{n}$ belongs to the first~(\emph{resp.} the second) subset if $a_{n}> b_{n}$ (\emph{resp.}~$a_{n}\leq b_{n}$). Choose the size of the first subset as~$p$ and the corresponding indices of terms in the first subset as~$l_1,l_2,\ldots,l_p$. It is clear that the chosen variables constitute an optimal solution to the optimization problem in this lemma. Thus, we see the optimum of the optimization problem is equal to~$\sum\nolimits_{n=1}^{N}[a_n-b_n]^+$.
\end{pf}
\begin{pf*}{PROOF OF THEOREM~\ref{tensorone}.}
This proof consists of two parts. In the first part, we transform the matrix completion problem into a maximum flow problem in an associated $s$-$t$ network, following the procedures in~\cite{chen2016constrained,fulkerson1959network}. In the second part, we show that there exists an integral $s$-$t$ flow with the desired value if and only if the defined structure tensor is nonnegative.
\begin{align*}
  1)~\textsl{Construct an Associated $s$-$t$ Network}~~~~~~~~~~~~~~~~~~~~~~~~~~~~~~~~~~~~~
\end{align*}
On account of the constraint (\ref{c3}), we can construct a directed bipartite graph $\mathcal{G}=\left(\mathcal{V}=\mathcal{V}_h\cup \mathcal{V}_r,\mathcal{E}\right)$, where
  \begin{alignat*}{8}
    &\mathcal{V}_h&{}={}&\left\{v_{h_1},v_{h_2},\cdots,v_{h_T}\right\},\\
    &\mathcal{V}_r&{}={}&\left\{v_{r_1}, v_{r_2},\cdots, v_{r_N}\right\},\\
    &\mathcal{E}&{}={}&\left\{(v_{h_j},v_{r_n})\ |\ F(n,j)=1\right\}.
  \end{alignat*}
Further, we construct an $s$-$t$ network based on~$\mathcal{G}$. Add a source node~$s$ and connect it to every node in $\mathcal{V}_h$; also, add a sink node $t$ and connect every node in $\mathcal{V}_r$ to $t$. The vertices in $\mathcal{V}$ now become internal nodes. Let~$\mathcal{\tilde{G}}=(\tilde{\mathcal{V}},\tilde{\mathcal{E}})$, where $\tilde{\mathcal{V}}=\mathcal{V}\cup\{s,t\}$ and $\tilde{\mathcal{E}}$ is the union of the arc set~$\mathcal{E}$ and the added arcs with one end as~$s$ or~$t$. In view of the constraints (\ref{c0})--(\ref{c2}), define the arc capacities on~$\tilde{\mathcal{E}}$ as
\begin{figure}[b]
  \begin{minipage}[t]{0.3\linewidth}
    \centering
   \mbox{ $\begin{bmatrix}
       1 & 1 & 1 & 1 & 0 & 0 \\
       1 & 1 & 1 & 1 & 0 & 0 \\
       1 & 1 & 1 & 1 & 1 & 1 \\
       0 & 0 & 0 & 1 & 1 & 1 \\
       0 & 0 & 0 & 1 & 1 & 1
     \end{bmatrix}$}
  \end{minipage} \vline
  \begin{minipage}[h]{0.68\linewidth}
    \centering
    \includegraphics[scale=0.62]{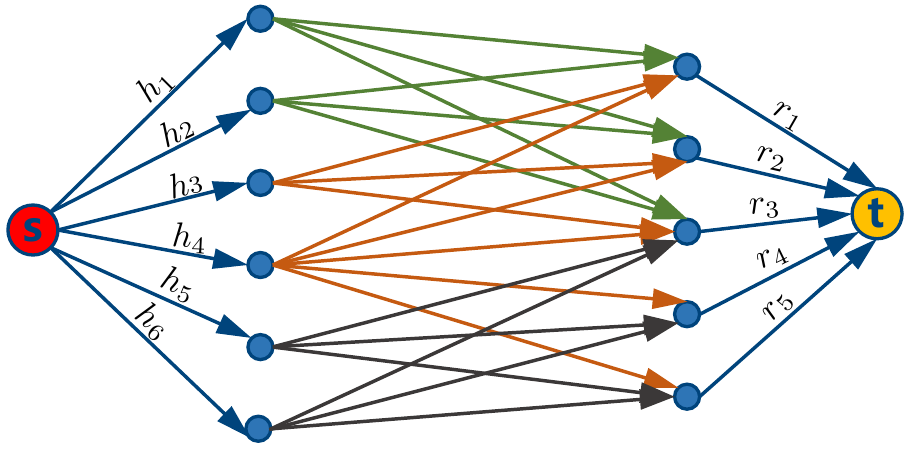}
  \end{minipage}
  \caption{A pattern matrix and its associated $s$-$t$ network.}
  \label{fig: flowgraph}
\end{figure}
  \begin{alignat*}{8}
    &c\left(s,v_{h_j}\right)=h_j, \text{~for every~} j=1,2,\dots,T;\\
    &c\left(v_{r_n},t\right)=r_n, \text{~for every~} n=1,2,\dots,N;\\
    &c\left(v_{h_j},v_{r_n}\right)=1, \text{~for every~} \left(v_{h_j},v_{r_n}\right)\in \mathcal{E}.
  \end{alignat*}
  The resulting~$s$-$t$ flow network~$(\mathcal{\tilde{G}},s,t,c)$ fully encapsulates the information underlying the matrix completion problem involving $\bm h$, $\bm r$ and $F$, as exemplified in~Fig.~\ref{fig: flowgraph}.

Observe that every matrix $A$ in $\mathcal{A}(\bm{h},\bm{r},F)$ can be associated with an integral flow~$f$ over~$\tilde{\mathcal{E}}$ in the $s$-$t$ network~$(\mathcal{\tilde{G}},s,t,c)$ with
\begin{alignat*}{8}
  &f\left(v_{h_j},v_{r_n}\right)=1, \text{~if~} A\left(n,j\right)=1;\\
  &f\left(v_{h_j},v_{r_n}\right)=0, \text{~if~} A\left(n,j\right)=0;\\
    &f\left(s,v_{h_j}\right)=\sum\nolimits_{u}f\left(v_{h_j},u\right), \text{~for every~} j=1,2,\dots,T;\\
    &f\left(v_{r_n},t\right)=\sum\nolimits_{u}f\left(u,v_{r_n}\right), \text{~for every~} n=1,2,\dots,N.
\end{alignat*}
Conversely, we can also associate every integral flow in the $s$-$t$ network with an $N\times T$ $(0,1)$-matrix belonging to~$\mathcal{A}(\bm{h},\bm{r},F)$. In addition, the value of such a flow~$f$ is~$$\left|f\right|=\sum_{j=1}^{T}f\left(s,v_{h_j}\right)=\sum_{n=1}^{N}f\left(v_{r_n},t\right).$$
It follows that finding a matrix in $\mathcal{A}(\bm{h},\bm{r},F)$ is equivalent to finding an integral flow of value~$\|\bm r\|_1$ in the associated $s$-$t$ network. Thus, we conclude by the Integral Flow theorem that the $(0,1)$-matrix class $\mathcal{A}(\bm h,\bm r ,F)$ is nonempty if and only if the maximum value of a flow in the associated $s$-$t$ network is $\|\bm r\|_1$ and $\|\bm h\|_1=\|\bm r\|_1$.
\begin{align*}
  2)~\textsl{Derive the Structure Tensor}~W(\bm h,\bm r, F)~~~~~~~~~~~~~~~~~~~~~~~~~~~~~~~~~~~~~
\end{align*}
In the first part, we have transformed the matrix completion problem into a maximum flow problem. By the Max-Flow-Min-Cut theorem, we can conclude that the matrix class $\mathcal{A}(\bm{h},\bm{r},F)$ is nonempty if and only if none of the $s$-$t$ cuts in the flow network~$(\mathcal{\tilde{G}},s,t,c)$ has a capacity less than $\|\bm r\|_1$ and $\|\bm h\|_1=\|\bm r\|_1$.

However, checking the capacity of every $s$-$t$ cut is unwise, because of an exponential number ($2^{N+T}$) of such cuts. The main idea of most methods involving inequalities to verify the non-emptiness of~$\mathcal{A}(\bm h,\bm r, F)$ is to exclude a large number of redundant calculations on these cuts. Next, we will clarify how to do this, which differs from other methods including that in our previous paper~\cite{chen2016constrained}.

Consider an arbitrary $s$-$t$ cut $(\mathcal{V}_s,\tilde{\mathcal{V}}\backslash\mathcal{V}_s)$. To be precise, let
$$\mathcal{V}_s=\{s\}\cup\{v_{h_j},j\in \mathcal{M}\}\cup\{v_{r_n},n\in\mathcal{\bar{L}}\},$$
where $\mathcal{M}\subseteq \mathcal{V}_h, \mathcal{L}\subseteq \mathcal{V}_r$ and $\mathcal{\bar{M}}=\mathcal{V}_h\backslash\mathcal{M}, \mathcal{\bar{L}}=\mathcal{V}_r\backslash\mathcal{L}$. 
Accordingly, the complementary set $$\tilde{\mathcal{V}}\backslash\mathcal{V}_s=\{t\}\cup\{v_{h_j},j\in \mathcal{\bar{M}}\}\cup\{v_{r_n},n\in \mathcal{L}\}.$$ Note that the size of $\mathcal{M}$ ranges from $0$ to $T$, while the size of~$\mathcal{L}$ has a range from $0$ to $N$. Let 
\begin{align*}
\mathcal{M}&~=~\left\{v_{h_{m_1}},v_{h_{m_2}},\cdots,v_{h_{m_q}}\right\},\\
\mathcal{L}&~=~\left\{v_{r_{l_1}},v_{r_{l_2}},\cdots,v_{r_{l_p}}\right\}.
\end{align*}
As demonstrated in Fig.~\ref{fig:cutsthreeparts}, the capacity $c(\mathcal{V}_s,\tilde{\mathcal{V}}\backslash\mathcal{V}_s)$ of the cut~$(\mathcal{V}_s,\tilde{\mathcal{V}}\backslash\mathcal{V}_s)$ with the prescribed $\mathcal{M}$ and $\mathcal{L}$ above comes from three elements: the arcs between the source~$s$ and vertices in $\mathcal{\bar{M}}$, the arcs between vertices in $\mathcal{\bar{L}}$ and the sink~$t$, and the arcs directed from vertices in $\mathcal{M}$ to vertices in $\mathcal{L}$. Since~$c(\mathcal{V}_s,\tilde{\mathcal{V}}\backslash\mathcal{V}_s)$ is completely determined by~$(\mathcal{M},\mathcal{L})$ and it is useful to have this dependence explicitly stated, let $\hat{c}$ be such that $\hat{c}(\mathcal{M},\mathcal{L}) = c(\mathcal{V}_s,\tilde{\mathcal{V}}\backslash\mathcal{V}_s)$. It follows that $\hat{c}(\mathcal{M},\mathcal{L})$ can be written as
\begin{equation*}
  \begin{split}
    \hat{c}(\mathcal{M},\mathcal{L})=&\left(\sum_{j=1}^Th_j-\!\!\sum_{j=1}^qh_{m_j}\right)+\left(\sum_{n=1}^Nr_n-\!\!\sum_{n=1}^pr_{l_n}\right)\\
&\ \ \ \ \ +\left(\sum_{n=1}^p\sum_{j=1}^qc(v_{h_{m_j}},v_{r_{l_n}})\right)
\end{split}
\end{equation*}
\begin{equation*}
\begin{split}
=&\left(\sum_{j=1}^Th_j-\sum_{j=1}^qh_{m_j}+\sum_{n=1}^Nr_n\right)\\
    \ \ &-\sum_{n=1}^p\left(r_{l_n}-\sum_{j=1}^qF(l_n,m_j)\right).
  \end{split}
\end{equation*}
As we can see,~$\hat{c}(\mathcal{M},\mathcal{L})$ can be written as the difference of two terms. Observe that the first term in~$\hat{c}(\mathcal{M},\mathcal{L})$ is only dependent on~$\mathcal{M}$. Using Lemma~\ref{keylem} allows us to explicitly optimize the second term over~$\mathcal{L}$. 
As a result, we can reduce the optimization problem $\min_{\mathcal{M},\mathcal{L}}\hat{c}(\mathcal{M},\mathcal{L})$ to $\min_\mathcal{M} w(\mathcal{M})$, where
\begin{figure}[t]
   \centering
   \includegraphics[scale=0.38]{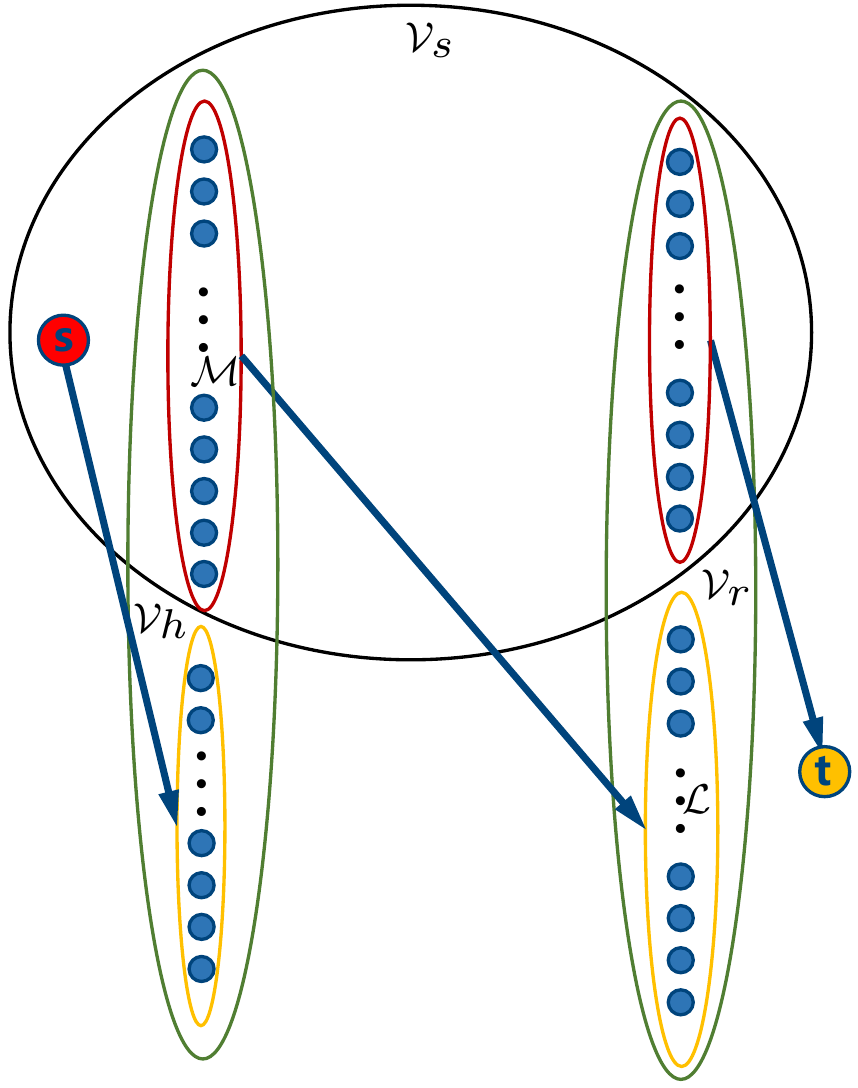} 
   \caption{Illustration of the $s$-$t$ cut $(\mathcal{V}_s,\tilde{\mathcal{V}}\backslash\mathcal{V}_s)$.}
   \label{fig:cutsthreeparts}  %
\end{figure}
\begin{alignat*}{8}
  w(\mathcal{M})&=\min_{\mathcal{L}}\hat{c}(\mathcal{M},\mathcal{L})\\
  &=\left(\sum_{j=1}^Th_j-\sum_{i=1}^qh_{m_i}+\sum_{n=1}^Nr_n\right)\\
  &\ \ \ \ -\sum_{n=1}^N\left[r_{n}-\sum_{i=1}^qF(n,m_i)\right]^+.
\end{alignat*}
Furthermore, we can conclude that the class $\mathcal{A}(\bm{h},\bm{r},F)$ is nonempty if and only if $w(\mathcal{M})\geq \|\bm r\|_1,\ \forall \mathcal{M}\subseteq \mathcal{V}_h$.

Next, we will study how to associate $w(\mathcal{M})\geq \|\bm r\|_1$ with the nonnegativity of the structure tensor $W(\bm h,\bm r, F)$ by optimizing $w(\mathcal{M})$ over $\mathcal{M}$.

Considering a specific~$\mathcal{{M}}$, if we can find two indices~$\hat{t}$ and $\tilde{t}$ such that
\begin{multline*}
  v_{h_{\hat{t}}}\notin \mathcal{M},v_{h_{\tilde{t}}}\in \mathcal{M}\text{, and} \\
  T_{i-1}<\hat{t}<\tilde{t}\leq T_{i}\text{, for a certain }i\in\{1,2,\ldots,\lambda\},
\end{multline*}then we can obtain a new vertex set~$\mathcal{\hat{M}}$ from $\mathcal{M}$ by replacing the vertex~$v_{h_{\tilde{t}}}$ with~$v_{h_{\hat{t}}}$. Furthermore, since $F$ is described by the formula~(\ref{despattern}) and $\bm h$ has the monotonicity assumption~(\ref{monoto}), we can verify~$w(\mathcal{\hat{M}})\leq w(\mathcal{M})$ easily.

Continue the search for such indices~$\hat{t}$ and $\tilde{t}$ to obtain a new vertex set until no such replacement can be conducted anymore. By the same argument, we see that the values of $w$ over these vertex sets are non-increasing in accordance with the order they appear. As a result, we can conclude that $\min_\mathcal{M} w(\mathcal{M})$ is equal to one of the following expressions
\begin{multline*}
\sum_{j>k_1}^{T_1}h_j+\sum_{j>T_1+k_2}^{T_2}h_j+\cdots+\sum_{j>T_{\lambda-1}+k_{\lambda}}^{T_{\lambda}}h_j+\sum_{n=1}^Nr_n\\
  -\!\sum_{n=1}^{N}\!\left[r_n\!\!-\!\!\sum_{i=1}^{\lambda}(F(n,T_{i-1}\!+1)+\cdots+F(n,T_{i-1}\!+k_i))\!\right]^+\!\!,
\end{multline*}
where~$k_i=0,1,\dots,T_i-T_{i-1}$, for $i=1,2,\dots,\lambda$. Comparing the above expression with the definition of the structure tensor~$W(\bm{h},\bm{r},F)$, we ultimately conclude that~$W(\bm{h},\bm{r},F)\geq 0$ if and only if $w(\mathcal{M})\geq \|\bm r\|_1$ for every subset $\mathcal{M}$ of $\mathcal{V}_h$, which completes the proof.
\end{pf*}
Theorem~\ref{tensorone} says that we can verify the non-emptiness of~$\mathcal{A}(\bm h, \bm r, E)$ with a first-order tensor. This is exactly in line with the expression~(\ref{OneOrderTensor}) which is closely related to the Gale-Ryser theorem. Moreover, when $F$ presents a staircase pattern, the tensor herein is consistent with the result obtained in our previous paper~\cite{chen2016constrained}. Hence, compared with these existing results, one of the contributions of Theorem~\ref{tensorone} is to provide a generic necessary and sufficient condition for the non-emptiness of~$\mathcal{A}(\bm h, \bm r, F)$ without requiring~special structures of~$F$. 

A physical interpretation of Theorem~\ref{tensorone} will be elaborated with an application to differentiated energy services in a later section. From the technical perspective, the proof of Theorem~\ref{tensorone} gives an equivalent network-flow formulation for the $(0,1)$-matrix completion problem considered in this paper. The matrix class~$\mathcal{A}(\bm h,\bm r, F)$ is not empty if and only if $\|\bm h\|_1=\| \bm r\|_1$ and the maximum flow value in an associated $s$-$t$ network is no less than the total sum of the row sum vector. A direct application of the Max-Flow-Min-Cut theorem requires us to check an exponential number~($2^{N+T}$) of inequalities, as shown by~\cite{fulkerson1959network,mirsky1968combinatorial}. By carefully applying several optimization techniques, we can remove a large number of calculations and finally derive a much simpler necessary and sufficient condition in the form of the nonnegativity of the structure tensor~$W(\bm{h},\bm{r},F)$. 

In addition, as we can see, the tensor condition makes full use of the description~(\ref{despattern}) for the pattern matrix~$F$, especially from the viewpoint regarding the tensor size, which is \mbox{$(T_1-T_0+1)\times(T_2-T_1+1)\times\!\cdots\!\times(T_\lambda-T_{\lambda-1}+1)$.} To a certain degree, this explains why we favor a pattern matrix~$F$ which can be described by a relatively small~$\lambda$ and has a rather large number of rows. In this case, the related structure tensor is of a manageable size and the tensor condition greatly reduces the computational complexity resulted from a large~$N$. Such a strength will be made more clear by the simulation results presented in the next subsection where we use the tensor condition to construct a matrix in~$\mathcal{A}(\bm h, \bm r, F)$.

\subsection{Complexity Analysis and Comparison}
As shown in the proof of Theorem~\ref{tensorone}, finding a matrix in~$\mathcal{A}(\bm h,\bm r, F)$ can be transformed into finding a flow of the maximum value in an associated $s$-$t$ network. Many well-known algorithms can be applied to finding such a maximal flow. Several classic flow algorithms include the Ford-Fulkerson algorithm and the Edmonds-Karp algorithm~\cite{edmonds1972theoretical,ford1956maximal}, whose complexities are respectively given by
$$\mathcal{O}\left(\|F\|_1\|\bm r\|_1\right)\text{ and }\mathcal{O}\left((N+T)\|F\|_1^2\right).$$
Both are based on the concept of path augmentation. Another class of algorithms features two basic operations, ``push'' and ``relabel'', firstly designed by Goldberg and Tarjan~\cite{goldberg1988new}. Such push-relabel algorithms generally have the complexity
$$\mathcal{O}\left((N+T)^2\|F\|_1\right).$$ Moreover, there are other kinds of flow algorithm, like the network simplex algorithms, which is a graphical specialization of the simplex algorithm~\cite{orlin1997polynomial}.

In view of Theorem~\ref{tensorone}, the complexity of calculating an element of $W(\bm{h},\bm{r},F)$ is given by~$\mathcal{O}(N+T)$. For the same~$T$ and $\lambda$, there are at most $(T/\lambda+1)^{\lambda}$ elements in this tensor. Therefore, the complexity of calculating~$W(\bm{h},\bm{r},F)$ is at most
$$\mathcal{O}\left((N+T)(T/\lambda+1)^{\lambda}\right).$$ We herein simply utilize the tensor condition as a checker to generate one matrix in~$\mathcal{A}(\bm h,\bm r, F)$, but hopefully we can design more subtle algorithms based on Theorem~\ref{tensorone} in the future. Specifically, for each unfixed position in a certain uncompleted column, we check whether there is a matrix in~$\mathcal{A}(\bm h,\bm r, F)$ with a one there by the tensor condition. If not, fix a zero at the position and go to the next unfixed position or the next uncompleted column when this column is fully filled. Therefore, the complexity of the tensor approach to finding a required matrix is at most~$$\mathcal{O}\left((N+T)(T/\lambda+1)^{\lambda}\|F\|_1\right).$$ As we can see, most existing network-flow algorithms are equally influenced by both row and column data, while the tensor approach treats rows and columns differently. In the following, we will show the efficiency of our tensor approach by numerical simulations for cases where $F$ has a
rather large number of rows and can be described by a relatively small $\lambda$.

Here is the setup: $\lambda=3$, $T_0=0$, $T_1=8$, $T_2=16$, and~$T_3=24$. Thus, there are three kinds of columns in the pattern matrix~$F$. For simplicity, we consider there are only three kinds of rows in~$F$ and the vectors corresponding to the three kinds of rows are:
\begin{figure}[t]
   \centering
\includegraphics[width=262pt,height=175pt]{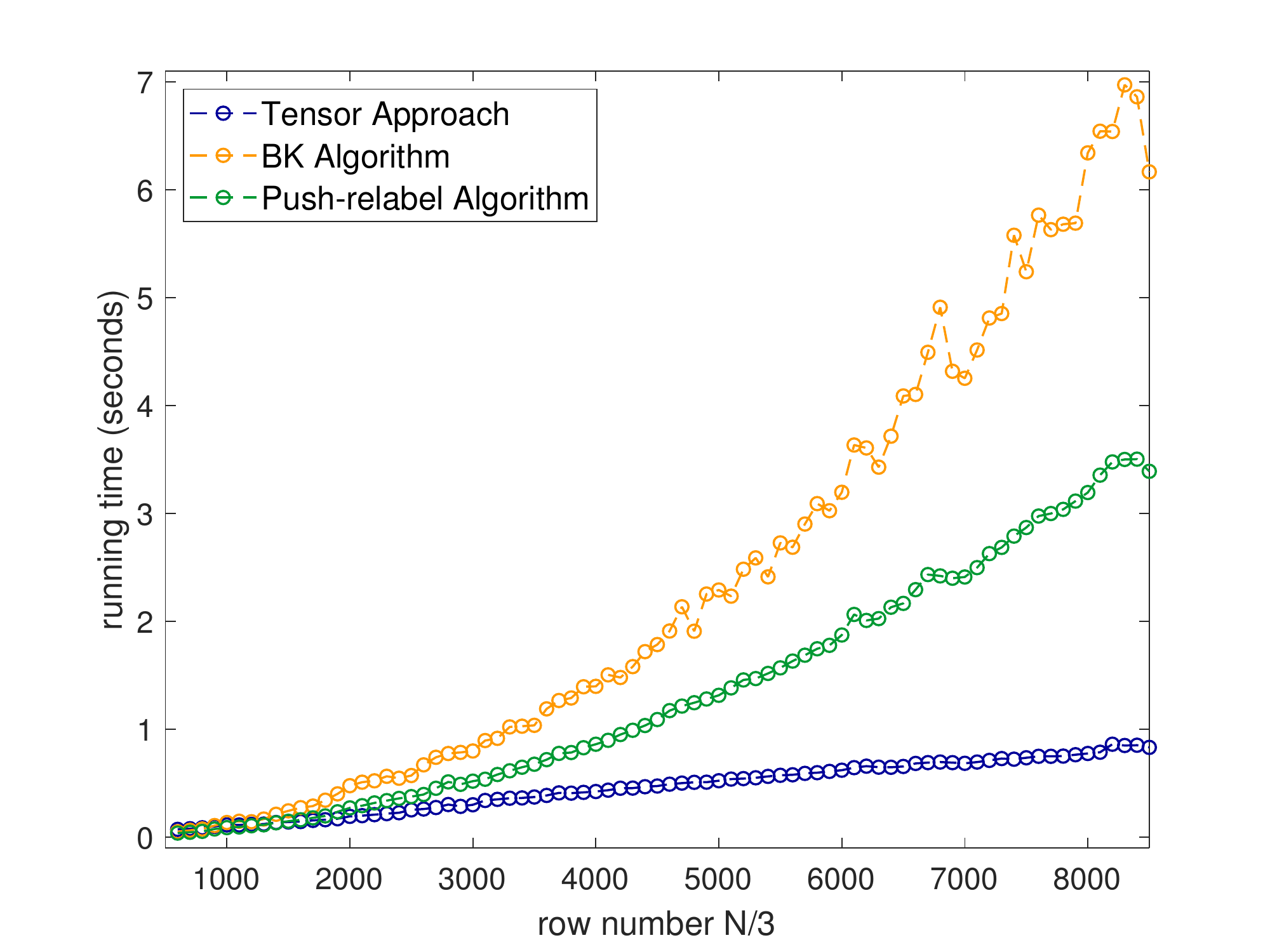} 
   \caption{Comparison between flow and tensor approaches.}
   \label{fig:complexitycomp}  %
\end{figure}
\begin{align*}
  &[1~1~1~1~1~1~1~1~1~1~1~1~1~1~1~1~0~0~0~0~0~0~0~0]'; \\
  &[1~1~1~1~1~1~1~1~1~1~1~1~1~1~1~1~1~1~1~1~1~1~1~1]'; \text{ and } \\
  &[0~0~0~0~0~0~0~0~1~1~1~1~1~1~1~1~0~0~0~0~0~0~0~0]'.
\end{align*}
Each kind has the same number of rows, which is $N/3$. Other relevant data, like~$\bm h$ and $\bm r$, are randomly generated from uniform distributions over their respective possible values. In terms of the running time for finding a matrix in~$\mathcal{A}(\bm h,\bm r, F)$, we compare the tensor approach with a default push-relabel algorithm in MATLAB and the Boykov-Kolmogorov~(BK) algorithm, which outperforms other augmenting-path flow algorithms in many experimental cases~\cite{boykov2004experimental}. We run each instance for twenty times and record the average running time. Following are observed phenomena.

\begin{figure}[t]
   \centering
\includegraphics[width=262pt,height=175pt]{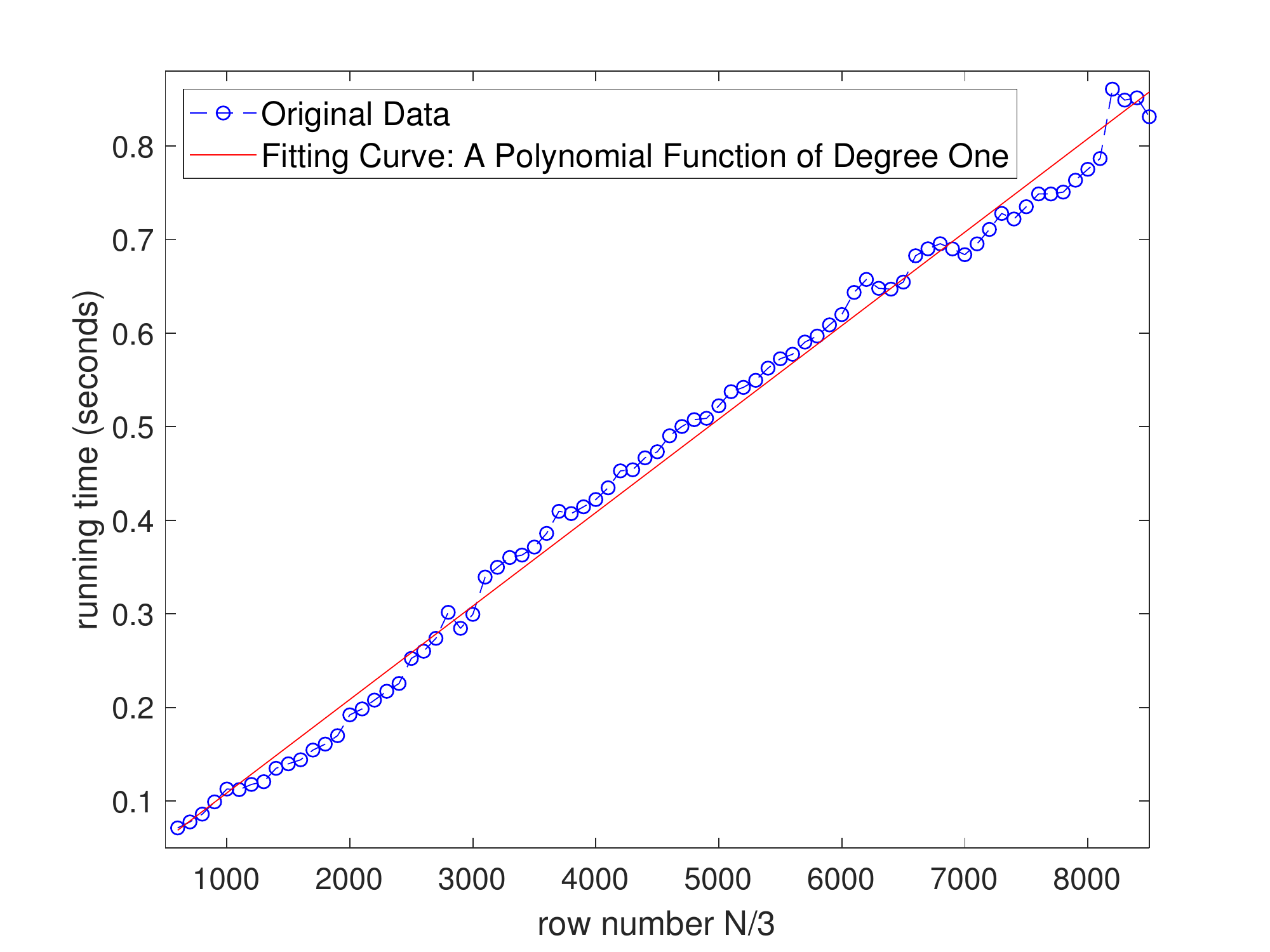} 
   \caption{Running time of the tensor approach as the row number increases.}
   \label{fig:complexitylinear}  %
\end{figure}

As depicted in~Fig.~\ref{fig:complexitycomp}, the strength of the tensor approach is more apparent as the row number increases. In this case, we also see that the push-relabel algorithm surpasses the BK algorithm. This is slightly different from the complexities stated. To be specific, only in terms of~$N$, the complexities of the tensor approach and the Ford-Fulkerson algorithm are approximately $\mathcal{O}\left(N^2\right)$ while the complexity of the push-relabel algorithm is approximately~$\mathcal{O}\left(N^3\right)$. In~Fig.~\ref{fig:complexitylinear}, we use a linear model to fit the tensor-approach data (in blue color) in~Fig.~\ref{fig:complexitycomp}. Numerically, we show that the complexity of the tensor approach grows linearly with respect to the row number, which is actually better than the aforementioned theoretical complexity. However, the increasing rates of other two algorithms are both superlinear with respect to the row number as visibly seen in~Fig.~\ref{fig:complexitycomp}. That explains why our tensor approach numerically favors the case where the pattern matrix has a large $N$ and can be described by a small $\lambda$.
\begin{figure}[t]
   \centering
\includegraphics[width=262pt,height=175pt]{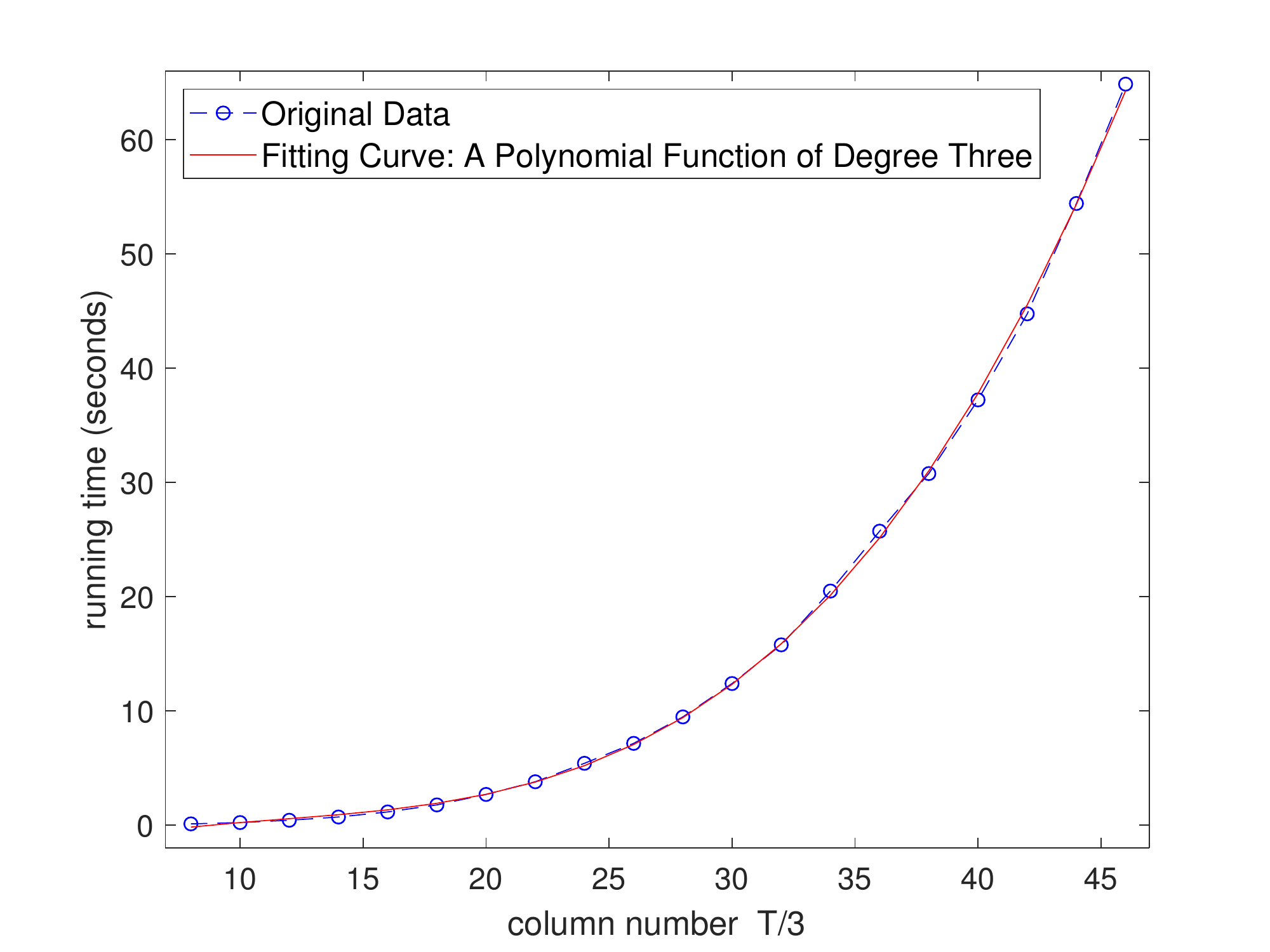} 
   \caption{Running time of the tensor approach as the column number increases.}
   \label{fig:complexitythree}  %
\end{figure}

Next, we fix the row number~$N/3=1000$ and increase the column number~$T/3$. It is shown in Fig.~\ref{fig:complexitythree} that the complexity curve of the tensor approach is fitted by a polynomial of degree three. Thus, we see that the complexity of our tensor approach grows linearly with respect to the number of elements in the associated tensor, which is~$(T/3+1)^3$ here. From this perspective, we conclude from simulation results that our tensor approach favors the case with a smaller tensor size. As a whole, we have numerically explained why our results are particularly useful when~$F$ has a rather large number of rows and can be described by a relatively small~$\lambda$.

\section{Differentiated Energy Services}
To better illustrate our results about the $(0,1)$-matrix completion, we apply them to solving allocation problems in differentiated energy services. 
As an implementation of demand response for future smart grid, electricity services herein are no longer homogeneous products sold at a common unit price, but a package of heterogeneous energy services differentiated by different levels of flexibility~\cite{nayyar2016duration,mo2017differentiated}. In this section, we address two adequacy problems in differentiated energy services, namely, the supply/demand matching and the minimum purchase profile problems. We begin with the model formulation.
\subsection{Model Formulation: Supply/Demand Matching}
Above all, we will elaborate how the supply and demand interact with each other in the so-called differentiated energy services. The connection between the adequacy problems in the model and the $(0,1)$-matrix completion studied in this paper will be made clear as we proceed.

Divide the operational period into sequential time slots, indexed by~$$j\in\{1,2,\dots,T\}.$$ At each time slot, a load can either be charged at a base power delivery rate or receive no power. The charging process during one time slot can never be suspended. In the next section, we will generalize the model by allowing involved loads to be charged at multiples of the base rate. Without loss of generality, we assume that the base power delivery rate is one unit per time slot and other quantities are scaled correspondingly. At the $j$th time slot, there are~$h_j$ units of electrical energy available from the supply. All the available units make up the \emph{supply profile}:
\begin{equation*}
  \bm{h}~=~\left[h_1~h_2~\cdots~h_T\right]'.
\end{equation*}
There are a collection of loads, indexed by $$n\in\{1,2,\dots,N\}.$$ Each load~$n$ has a duration requirement, denoted by~$r_n$, which indicates how many units of energy it requires.

The involved loads can only be charged during their service times. Firstly, the supplier provides $\lambda+1$ special time instances:
$$T_0<T_1<T_2<\cdots<T_{\lambda-1}<T_\lambda,$$
wherein $T_0=0$ and $T_\lambda=T$. Then, the service time of load~$n$ is specified by the pair~$\left(T_{a_n},T_{d_n}\right)$ which means the power delivery to load~$n$ is only able to happen from the~$\left(T_{a_n}+1\right)$th to the $T_{d_n}$th time slot; moreover, the service time vector of load~$n$ refers to the~$(0,1)$-vector of length~$T$ whose indices of ones exactly range from~$T_{a_n}+1$ to $T_{d_n}$. Overall, a load~$n$ or a differentiated energy service~$\left(r_n,T_{a_n},T_{d_n}\right)$ is specified by the duration requirement~$r_n$ and the service time~$\left(T_{a_n},T_{d_n}\right)$. Such a service is also called the multiple-arrival multiple-deadline differentiated energy service, primarily studied in our conference paper~\cite{mo2017differentiated}. If all the loads share the same service time, we recall a prototype of differentiated energy services called the duration-differentiated energy services~\cite{nayyar2016duration}, since the difference between two distinct services merely lies in the duration requirement. In addition, we refer to the case where all the loads have the same arrival time but diverse deadlines as the duration-deadline jointly differentiated energy services~\cite{chen2016constrained}.

Moreover, a feasible charging profile for a load is denoted by a~$(0,1)$-vector of length~$T$, whose sum is the duration requirement and nonzero elements only appear in positions indicated by the service time of the load.

To clarify the notation, we summarize all the duration requirements as the~\emph{demand profile}: $$\bm{r}~=~[r_1~r_2~\cdots~r_N]'.$$
In addition, all the service times form a pattern matrix~$F$ such that the service time vector of load~$n$ is~$F(n,:)'$. By the formula~(\ref{despattern}), this~$F$ can be described by $\lambda+1$ special columns indices corresponding to the prespecified special time indices, but the ones in each row of~$F$ are consecutive herein. Moreover, in the duration-differentiated energy services, we have $F=E$, while in the duration-deadline jointly differentiated energy service, the rows can be permuted in such a way that the pattern matrix~$F$ presents a staircase pattern. In brief, all the requirements of the collection of loads are summarized as~$(\bm r, F)$. 

The first question coming into mind is whether the supply, specified by the supply profile~$\bm h$, can satisfy the collection of loads whose requirements are embodied in the demand profile~$\bm r$ and the pattern matrix~$F$. If so, then we say that the supply~$\bm h$ is adequate for the demand~$(\bm r, F)$. Otherwise, it is inadequate. Furthermore, in the case of an adequate supply, there exists a feasible power allocation to redistribute the electrical energy over time from the supplier to the collection of loads, which corresponds to a matrix~$A$ satisfying the following constraints:
\begin{eqnarray} \label{c4}
  (\ref{c0});(\ref{c1}); (\ref{c3}); \nonumber \\
  ~~~~~~~~~~~~~~~~~~\left\|A(:,j)\right\|_1\leq h_j,\ \ \forall j=1,2,\dots,T.
\end{eqnarray}
Such a matrix~$A$ is called a feasible power allocation matrix since~$A(n,:)'$ can be a feasible charging profile for each load~$n$. Evidently, when $\|\bm h\|_1=\|\bm r\|_1$, the $(0,1)$-matrix class~$\mathcal{A}(\bm h,\bm r, F)$ is just the feasible power allocation matrix class specified by~$(\bm h, \bm r, F)$. The supply/demand matching problem is concerned with the conditions regarding~$(\bm h,\bm r,F)$ under which there exists a feasible power allocation matrix. In~\cite{nayyar2016duration} where~$F=E$, the problem has been solved as follows.
\begin{prop}\label{duration}
In the duration-differentiated energy services, the supply~$\bm h$ is adequate for the demand~$(\bm r, E)$ if and only if $\bm h\prec^w \bm r^*$, or equivalently $\bm r \prec_w \bm h^*$.
\end{prop}
This results is essentially the Gale-Ryser theorem without $\|\bm h\|_1=\|\bm r\|_1$. Thus, we expect to use our tensor condition to check the adequacy of a supply for a demand by analogy. Since there are consecutive ones in every row of the pattern matrix~$F$ derived from service times, we can represent the term~$\sum_{i=1}^{\lambda}k_iF(n,T_i)$ in the tensor defined in Section~$3$ by~$\left(k_{a_n+1}+k_{a_n+2}+\cdots+k_{d_n}\right)$; indeed, in the differentiated energy services, the $\lambda$th-order structure tensor is specially written as
\begin{equation*}
\begin{split}
  W_{k_1k_2\cdots k_{\lambda}}(\bm{h},\bm{r},F)\!= \! \! \sum_{j>k_1}^{T_1}\!h_j+\!\!\!\!\sum_{j>T_1+k_2}^{T_2}\!\!\!\!h_j+\cdots+\!\!\!\!\!\!\!\sum_{j>T_{\lambda-1}+k_{\lambda}}^{T_{\lambda}}\!\!\!\!h_j\\
  -\sum_{n=1}^{N}\left[r_n-\left(k_{a_n+1}+k_{a_n+2}+\cdots+k_{d_n}\right)\right]^+, \end{split}
\end{equation*}
where~$k_i=0,1,\dots,T_i-T_{i-1}$, for $i=1,2,\dots,\lambda$. The following theorem, which is specialization of~Theorem~\ref{tensorone} and firstly presented in the conference paper~\cite{mo2017differentiated} without proofs, addresses the supply/demand matching problem in the more general differentiated energy services.

\begin{thm}\label{tensortwo}
   The supply $\bm h$ is adequate for the demand~$(\bm r, F)$ if and only if $W(\bm h,\bm r,F)\geq 0$.
\end{thm}
With the aid of this engineering application, we elaborate the physical implication of the tensor condition. When $\lambda$ is one, the tensor is reduced to a vector of length~$T+1$: for $k=0,1,\dots,T$,
\begin{equation}\label{simpletails}
W_k(\bm h, \bm r,E)=\sum_{j>k}^Th_j - \sum_{n=1}^{N}[r_n-k]^+.
\end{equation}
The minuend of~(\ref{simpletails}) is obtained by summing up the least~$T-k$ elements of the supply profile and thus called the supply tail. Correspondingly, the subtrahend of
~(\ref{simpletails}) can be interpreted as the demand tail, which equals the summation of the least $T-k$ elements of the conjugate of the demand profile. As a result, the adequacy of the supply~$\bm h$ for the demand~$(\bm r,E)$ can be derived from the fact that the demand tail is always dominated by the supply tail for every critical point indexed by~$k=0,1,\dots,T$.

\begin{figure}[t]
\centering
\includegraphics[scale=0.52]{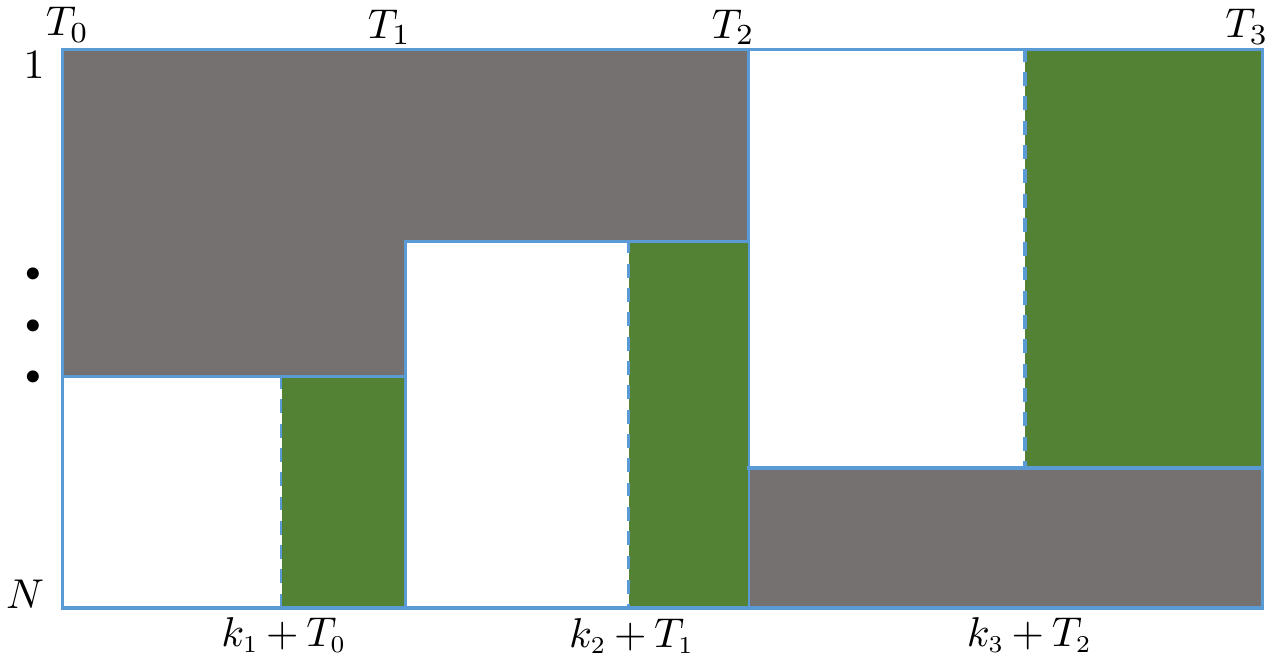}
   \caption{Illustration of the supply and demand tails.}
   \label{fig:tails}
\end{figure}

In a similar manner, such a tail dominating phenomenon can also be observed in the $\lambda$th-order structure tensor. For the supply tail, we refer to $$ \sum_{j>k_1}^{T_1}\!h_j+\!\sum_{j>T_1+k_2}^{T_2}\!h_j+\cdots+\!\sum_{j>T_{\lambda-1}+k_{\lambda}}^{T_{\lambda}}\!\!\!h_j.$$ In contrast, for the demand tail, we refer to $$\sum_{n=1}^{N}\left[r_n-\left(k_{a_n+1}+k_{a_n+2}+\cdots+k_{d_n}\right)\right]^+.$$
Each supply/demand tail pair is indexed by an index vector of length~$\lambda$, i.e., $[k_1~k_2~\cdots~k_\lambda]'$. Moreover, the tails herein are scattered. Take the $3$rd order tensor in Fig.~\ref{fig:tails} for illustration, where there are four special time instances~(namely, $T_0,T_1,T_2,T_3$) and grey areas correspond to fixed zeros. We classify loads into four groups by their service times, which are successively specified by~$\left(T_2,T_3\right),\left(T_1,T_3\right),\left(T_0,T_3\right)$ and $\left(T_0,T_2\right)$. For a fixed index vector~$[k_1~k_2~k_3]'$, we aggregate the colored part over the time horizon to get the supply tail. Accordingly, we accumulate the colored part over the load index to obtain the demand tail. Due to the pattern matrix~$F$ (different service times), the supply tail consists of three sub-tails in total, each of which is between two sequential special time instances. Likewise, the demand tail comes from the four groups of loads and each group has different sub-tails. For instance, the first load group has only one sub-tail, while the fourth group has two distinct sub-tails.

As a whole, the physical interpretation of our structure-tensor condition is as follows. The nonnegativity of each element in~$W(\bm h, \bm r, F)$ corresponds to the dominance relationship between an associated supply/demand tail pair. That is, the supply is adequate for the given demand if and only if the supply tail dominates the demand tail at every critical point indexed by $[k_1~k_2~\cdots~k_\lambda]'$, where~$k_i=0,1,\dots,T_i-T_{i-1}$, for $i=1,2,\dots,\lambda$.
\subsection{Adequacy Gap and Minimum Purchase Profile}
After addressing the supply/demand matching problem, a follow-up question arises -- what is the minimum supplementary purchase in case of inadequate supplies?
Let us define an auxiliary integer vector~$\bm p$ of the same length with the supply profile~$\bm h$ and call it the \emph{purchase profile}:
$$\bm{p}~=~\left[p_1~p_2~\cdots~p_T\right]'.$$
If the supply is inadequate, we are required to find a purchase profile~$\bm p$ such that the augmented supply profile~$\bm h+ \bm p$ is adequate and the total purchasing amount~$\|\bm p\|_1$ is as small as possible. If $\|\bm p\|_1$ is the minimum, then we call $\bm p$ a minimum purchase profile. Mathematically, the minimum purchase profile problem is formulated as
\begin{align}
\min_{\bm{p}}\quad& \|\bm p\|_1\tag{Minimum Purchase}\\
\text{s.t.}\quad& \bm p \in \mathbb{N}^T,\nonumber \\
&W(\bm h+\bm p,\bm r,F)\geq 0.\label{progap}
\end{align}
The optimal value of Problem~(\ref{progap}) is referred to as the \emph{adequacy gap}, which is the minimum amount of the additional purchase such that the augmented supply is adequate. Undoubtedly, when the supply is adequate, the adequacy gap is zero. The following theorem signifies that the adequacy gap can be obtained as a by-product of our tensor condition in Theorem~\ref{tensortwo}, whose proof follows directly from the derivation of the associated structure tensor $W(\bm h,\bm r, F)$. 
\begin{algorithm}[t]
\caption{Minimum Purchase Profile}\label{alg1}
\hspace*{0.02in}{\bf Input:}
An inadequate supply~$\bm h$ and a demand~$(\bm r, F)$;\\
\hspace*{0.02in}{\bf Output:}
A minimum purchase profile~$\bm p$.
\begin{algorithmic}[1]
\State {\bf Initialization:} $\bm p=O\in \mathbb{N}^T, \bm h^{1}=\bm h, j=1,v^o=0$, and $v^u=-\min_{k_1, k_2,\cdots, k_{\lambda}}W_{k_1k_2\cdots k_{\lambda}}(\bm{h},\bm r, F)$;
\For{$j\leq T$}
\State $v^o=v^u, \bm h^{j+1}=\bm h^{j}$;
\State $h^{j}_j=\min\left\{h^{j}_j+v^o,\|F(:,j)\|_1\right\}$;
\State $v^u=-\min_{k_1, k_2,\cdots, k_{\lambda}}W_{k_1k_2\cdots k_{\lambda}}(\bm{h}^{j},\bm r, F)$; \label{keystep}
\State $h^{j+1}_j=h^{j+1}_j+v^o-v^u,j=j+1$;
\If {$v^u= 0$} \State $\bm p=\bm h^j-\bm h$; \textbf{break};
\EndIf
\EndFor
\State {\bf Return:} A minimum purchase profile~$\bm p$.
\end{algorithmic}
\end{algorithm}
\begin{thm}\label{thmgap}
  The optimum of the minimum purchase profile problem~(\ref{progap}), i.e., the adequacy gap, is given by \\ \centering $\left |\min\limits_{k_1, k_2,\cdots, k_{\lambda}}W_{k_1k_2\cdots k_{\lambda}}(\bm h,\bm r,F)\right|.$
\end{thm}
Actually, the adequacy gap is given by the difference of the total demand and the maximum value of a flow in the associated $s$-$t$ network. We expect to eliminate the difference by designing a purchase profile of the minimum amount. We can easily show that such purchase profiles are not unique in general. 
Based on the intuitive algorithm presented in~\cite{mo2017differentiated}, we herein give two modified algorithms to find one of the minimum purchase profiles, by virtue of the structure tensor.

The correctness of Algorithm~\ref{alg1} follows directly from the proof of Theorem~\ref{tensorone}. The intuition of designing Algorithm~\ref{alg1} is straightforward. For $j=1,2,\ldots,T$ in the order stated, we repeatedly apply Theorem~\ref{thmgap} in Line~\ref{keystep} of the algorithm to compute at most how many units can be added to~$h_j$ without redundancy. It is possible to obtain a more efficient algorithm by exploring more interesting properties. This is Algorithm~\ref{alg2} described in what follows.

Before verifying the correctness of Algorithm~\ref{alg2}, we state two useful lemmas. The first lemma follows from the results in~\cite{mo2017coordinating}. Given~$\tau \in \mathbb{N}$ and~$\bm{\hat{h}}=\bm {\hat{h}}^{\downarrow}\in N^T$, we define~$$v=\min\left\{x\in \mathbb{N}~\middle|~\sum_{j=1}^{T}\left[x-{\hat{h}}_j\right]^+\geq \tau\right\}.$$ We derive a new vector $\bm {\tilde{h}}\in \mathbb{N}^T$ by firstly replacing every element of~$\bm{\hat{h}}$ which is less than~$v$ by $v$ and then decreasing each of the last $\sum_{j=1}^{T}[v-{\hat{h}}_j]^+-\tau$ elements by one.
\begin{algorithm}[t]
\caption{Minimum Purchase Profile} \label{alg2}
\hspace*{0.02in}{\bf Input:}
An inadequate supply~$\bm h$ and a demand~$(\bm r, F)$;\\
\hspace*{0.02in}{\bf Output:}
A minimum purchase profile~$\bm p$.
\begin{algorithmic}[1]
\State {\bf Initialization:} \mbox{$\bm p=O\!\in \!\mathbb{N}^T, \bm h^{0}=\bm h,i=k=v=0$}, \mbox{$v^o=0$, and $v^n\!=\!-\min_{k_1, k_2,\cdots, k_{\lambda}}\!W_{k_1k_2\cdots k_{\lambda}}(\bm{h},\bm{r},F)$;}
\For{$i<\lambda$}
\State $v^o=v^n$;
\State $v=\min\{x\in \mathbb{N}~|~\sum_{j=T_i+1}^{T_{i+1}}[x-h^{i}_j]^+\geq v^o\}$;
\State $h^{i+1}_j=\max \{ h^{i}_j, v\}$, if $T_i+1\leq j\leq T_{i+1}$;
\Statex \hspace{1.2em} otherwise, $h^{i+1}_j=h^{i}_j$;
\State $v^n=-\min_{k_1, k_2,\cdots, k_{\lambda}}W_{k_1k_2\cdots k_{\lambda}}(\bm{h}^{i+1},\bm{r},F)$;
\If{$v^n \leq 0$}
\State $k=\sum_{j=T_i+1}^{T_{i+1}}[v-h_j]^+\!-v^o, \bm q=\bm h^{i+1}-\bm h^i$;
\State Randomly pick exactly $k$ nonzero elements
\Statex \hspace{2.7em} in~$\bm q$ and decrease them by one;
\State $\bm p = \bm h^i + \bm q - \bm h$; {\bf break};
\Else
\State  $v\!\!=\!\!\max\{x\in \mathbb{N}~|~\sum_{j=T_i+1}^{T_{i+1}}[x-h^{i}_j]^+ \!\!\leq\! (v^o\!-\!v^n)\}$;
\State $h^{i+1}_j=\max \{ h^{i}_j, v\}$, if $T_i+1\leq j\leq T_{i+1}$;
\Statex \hspace{2.7em} otherwise, $h^{i+1}_j=h^{i}_j$;
\State $k=(v^o-v^n)-\sum_{j=T_i+1}^{T_{i+1}}[v-h_j]^+$;
\State Randomly pick $k$ elements and increase them
\Statex \hspace{2.77em} by one in $\{h^{i+1}_j~|h^{i+1}_j=v, T_i+1\leq j\leq T_{i+1}\}$;
\State $i=i+1$;
\EndIf
\EndFor
\State {\bf Return:} A minimum purchase profile~$\bm p$.
\end{algorithmic}
\end{algorithm}
\begin{lem}[Valley-filling~\cite{mo2017coordinating}]~\label{valleyfill}
The vector~$\bm{\tilde{h}}$ constructed above is majorized by every element in the set $\big\{\bm{\hat{h}}+\bm b~|~\bm b\in \mathbb{N}^T, \|\bm b\|_1=\tau\big\}$.
\end{lem}
The second lemma presented below follows from the Gale-Ryser theorem and Proposition~\ref{duration}.
\begin{lem} \label{colone}
  If the matrix class~$\mathcal{A}(\bm{\hat{h}}, \bm r, E)$ is nonempty and $\bm{\tilde{h}}\prec \bm{\hat{h}}$, then the matrix class~$\mathcal{A}(\bm{\tilde{h}}, \bm r, E)$ is nonempty.
\end{lem}
By Lemma~\ref{colone}, if we replace the sub-supply profile between any two sequential special time instances, i.e., $$\left[h_{T_i+1}~h_{T_i+2}\cdots~h_{T_{i+1}}\right]', i=0,1,\dots,\lambda-1,$$ with another nonnegative integer vector which is smaller in the majorization order, then the new supply is more adequate for the same demand~$(\bm r,F)$ than the old one in the sense that the new feasible power allocation matrix class is nonempty as long as the old one is nonempty.

\begin{pf*}{Correctness of Algorithm~\ref{alg2}}
In each iteration of Algorithm~\ref{alg2}, we figure out at most how much the adequacy gap can be reduced by properly augmenting the corresponding sub-supply profile~$[h_{T_i+1}~h_{T_i+2}\cdots~h_{T_{i+1}}]'$, where $i\in\{0,1,\dots,\lambda-1\}$. According to Lemma~\ref{valleyfill} and Lemma~\ref{colone}, the new supply profile in each iteration indeed decreases the adequacy gap by the corresponding maximum amount without redundancy. Furthermore, by the proof of Theorem~\ref{tensorone}, we can conclude that the adequacy gap will vanish after at most~$\lambda$ such iterations regarding different sub-supply profiles. Thus, Algorithm~\ref{alg2} generates a minimum purchase profile.
\end{pf*}
\begin{figure}[t]
   \centering
\includegraphics[width=262pt,height=183pt]{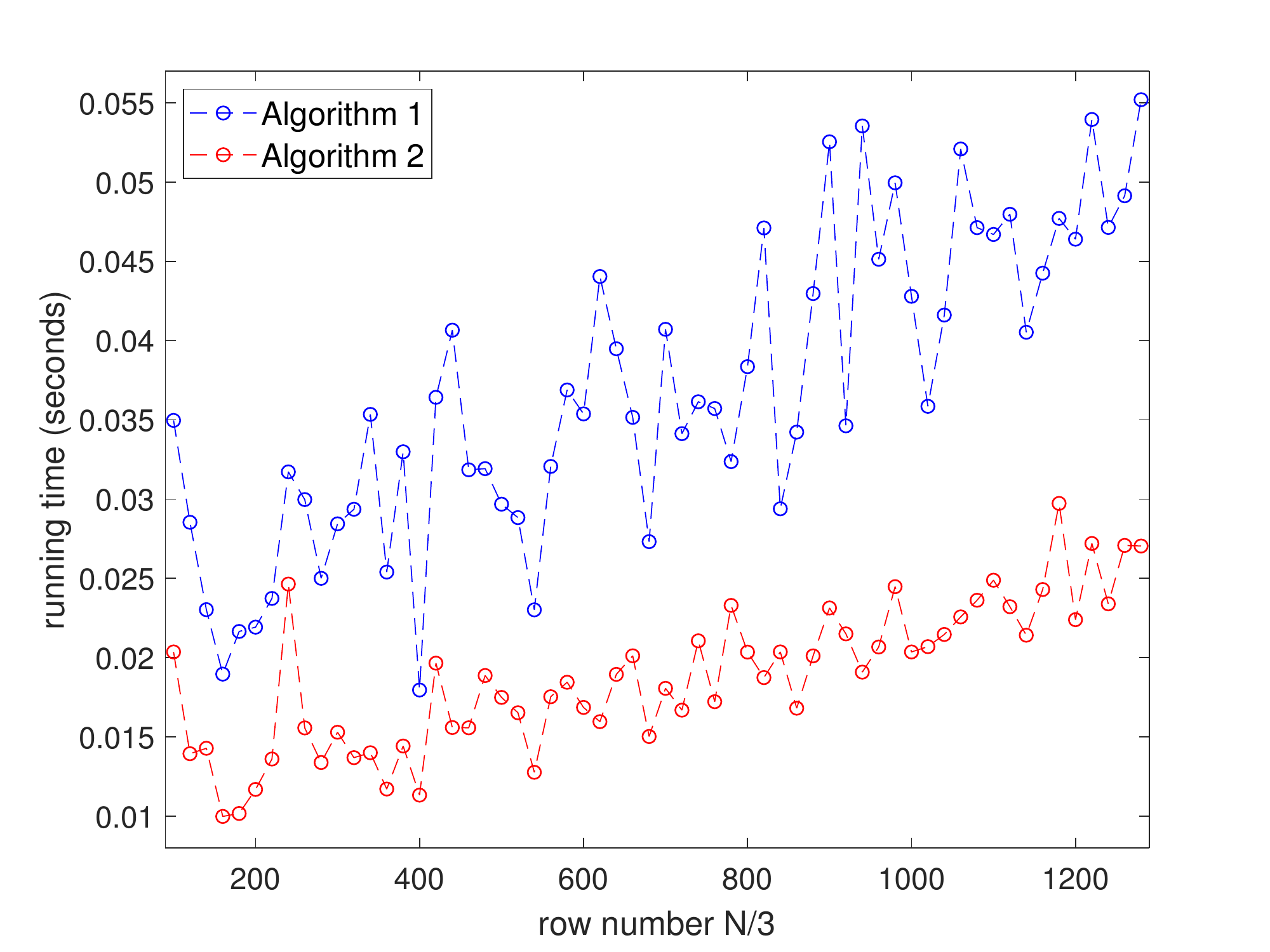} 
   \caption{Comparison between Algorithm~\ref{alg1} and Algorithm~\ref{alg2} for different row numbers.}
   \label{fig:compN}  %
\end{figure}
Both algorithms obtain a minimum purchase profile in a sequential way and the complexity of each iteration mainly comes from calculating the current tensor. However, in each iteration, Algorithm~\ref{alg2} deals with a bundle of columns indexed by labels between two sequential special time instances rather than a single column as Algorithm~\ref{alg1} does. Hence, the maximum number of iterations is reduced from the column number~$T$ in Algorithm~\ref{alg1} to the tensor order~$\lambda$ in Algorithm~\ref{alg2}. Overall, Algorithm~\ref{alg1} is easier to implement, while Algorithm~\ref{alg2} is more efficient because it does not only increase the supply amount but also further improves the distribution of the total supply in terms of the majorization order.

\begin{figure}[t]
   \centering
\includegraphics[width=262pt,height=183pt]{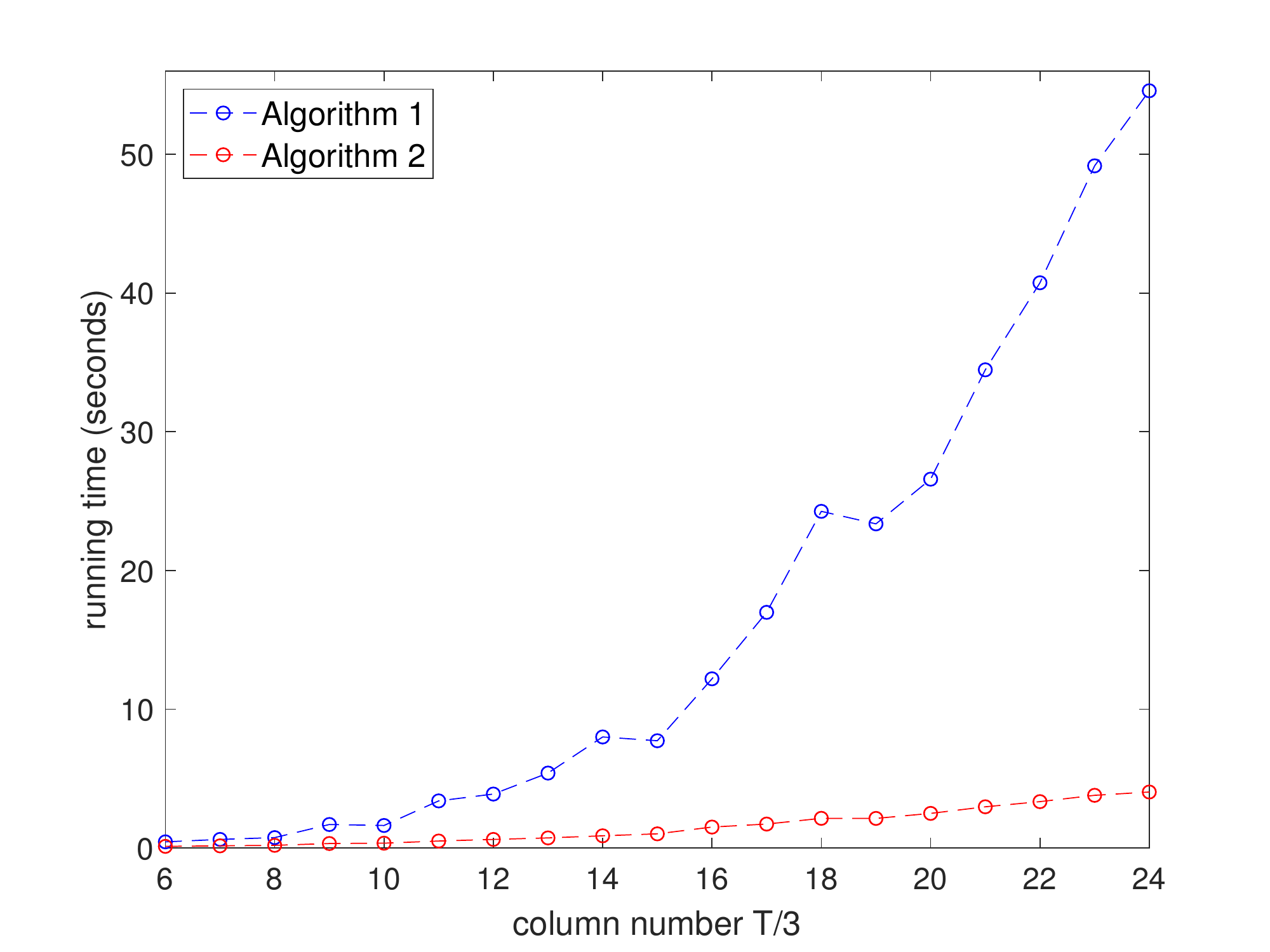} 
   \caption{Comparison between Algorithm~\ref{alg1} and Algorithm~\ref{alg2} for different column numbers.}
   \label{fig:compT}  %
\end{figure}

Numerical comparisons are given in Fig.~\ref{fig:compN} and Fig.~\ref{fig:compT}. We adopt the setup in Section~$3$, where $\lambda=3$, $T_3-T_2=T_2-T_1=T_1-T_0=T/3$ and $F$ has three kinds of rows. In Fig.~\ref{fig:compN}, we fix $T/3=8$ and compare the running times of two algorithms for different row numbers~$N/3$. In Fig.~\ref{fig:compT}, we fix $N/3=600$ and compare the running times of two algorithms for different column numbers~$T/3$. From the simulation results, we see that Algorithm~\ref{alg2} outperforms Algorithm~\ref{alg1} in terms of the running time. In addition, such an advantage is increasingly obvious as we increase the row/column number. However, the running time difference is more sensitive to the change of the column number than that of the row number, which is consistent with our previous theoretical analysis.

\section{Rate-constrained Differentiated Energy Services and Integer Matrix Completion}
After studying the two adequacy problems for the basic differentiated energy services, in this section, we move forward with more complicated rate-constrained differentiated energy services. In this framework, we revisit the problems of supply/demand matching and minimum purchase profile, wherein more general nonnegative integer matrix completion problems are considered. They involve the class of nonnegative integer matrices with prescribed row/column sums, predetermined zeros, and different bounds across the rows. Based on the results derived in the previous sections, we will address these problems with the help of a generalized structure tensor.

In a practical setup, different loads may not have a uniform power delivery rate as in the last section, which motivates us to study rate-constrained differentiated energy services~\cite{negrete2016rate}. The supply/demand model herein is almost the same as before. The difference lies in that, at each time slot, load~$n$ in this model can either receive no power or be charged at a certain multiple of the base power delivery rate. This multiple can be any integer in the interval~$[1,\bar{r}_n]$, where~$\bar{r}_n$ is called the ceiling charging rate of load~$n$. Overall, a load~$n$ or a rate-constrained differentiated energy service~$\left(r_n,\bar{r}_n,T_{a_n},T_{d_n}\right)$ is specified by the duration requirement~$r_n$, the ceiling charging rate~$\bar{r}_n$, and the service time~$\left(T_{a_n},T_{d_n}\right)$.

Given a collection of loads indexed by $n\in\{1,2,\ldots,N\}$, we summarize all the ceiling charge rates as the \emph{ceiling rate profile}~$\bm{\bar{r}}\in \mathbb{N}^N$, whose $n$th element~$\bar{r}_n$ represents the ceiling charging rate of load~$n$. 
Let~$\bm r$ and~$F$ respectively be the demand profile and the pattern matrix determined by the service times of involved~$N$ loads. Thus, all the requirements can be summarized as~$(\bm r, \bm{\bar{r}},F)$.

Clearly, given an adequate supply profile~$\bm h$, there exists a one-to-one correspondence between a feasible power allocation and an~$N\times T$ integer matrix~$A$ satisfying the following constraints:
\begin{eqnarray*}
0\leq A(n,j)\leq \bar r_n,\ \forall n=1,2,\dots,N~\&~j=1,2,\dots,T;\\
  (\ref{c1});(\ref{c4})\text{; and }\mathds{1}(A) \leq F.
\end{eqnarray*}
Such a matrix~$A$ is also called a feasible power allocation matrix. Let~$\mathcal{A}(\bm h,\bm r, \bm{\bar{r}},F)$ denote the class of feasible power allocation matrices specified by~$(\bm h,\bm r, \bm{\bar{r}},F)$. Mathematically, the supply/demand matching problem herein asks under what conditions the nonnegative integer matrix class~$\mathcal{A}(\bm h,\bm r, \bm{\bar{r}},F)$ is nonempty. To answer this question, we define a modified structure tensor~$W(\bm h,\bm r, \bm{\bar{r}},F)$ by analogy with $W(\bm h,\bm r, F)$:
\begin{multline*}
\!\!\!\!\!\!W_{k_1k_2\cdots k_{\lambda}}(\bm{h},\bm r, \bm{\bar{r}},F)=\!\! \sum_{j>k_1}^{T_1}\!h_j+\!\!\!\!\sum_{j>T_1+k_2}^{T_2}\!\!\!h_j+\cdots+\!\!\!\!\!\!\sum_{j>T_{\lambda-1}+k_{\lambda}}^{T_{\lambda}}\!\!\!\!\!\!h_j\\
   -\sum_{n=1}^{N}\left[r_n-\bar{r}_n\left(k_{a_n+1}+k_{a_n+2}+\cdots+k_{d_n}\right)\right]^+,
\end{multline*}
where~$k_i=0,1,\dots,T_i-T_{i-1}$, for $i=1,2,\dots,\lambda$. As a generalization of Theorem~\ref{tensorone} and/or Theorem~\ref{tensortwo}, the following theorem gives a necessary and sufficient condition for the non-emptiness of~$\mathcal{A}(\bm h,\bm r, \bm{\bar{r}},F)$, and also addresses the supply/demand matching problem in the rate-constrained differentiated energy services by the structure tensor defined above. 
\begin{thm}\label{tensortwoceil}
   The supply $\bm h$ is adequate for the demand~$(\bm r, \bm{\bar{r}},F)$, or equivalently, the class~$\mathcal{A}(\bm h,\bm r, \bm{\bar{r}},F)$ is nonempty if and only if $W(\bm h,\bm r, \bm{\bar{r}},F)\geq 0$.
\end{thm}
Before proceeding, we introduce a lemma which is critical to the understanding of Theorem~\ref{tensortwoceil}. 
\begin{lem}\label{decomp}
Consider nonnegative integers $r$, $\bar{r}$, $p$, and~$q$ such that $p <\bar{r}$ and $r=q\bar{r}+p$. The following holds:
\begin{equation*}
  [r-k\bar{r}]^+=p[q+1-k]^++(\bar{r}-p)[q-k]^+, \ \forall~k\in \mathbb{N}.
\end{equation*}
\end{lem}
\begin{pf}
  Consider a vector of length~$\bar{r}$, with $p$ elements as $q+1$ and the remaining $\bar{r}-p$ elements as $q$. The lemma follows from the observation that both sides of the equation are equal to the same reverse leading partial sum of the partition conjugate of the considered vector.
\end{pf}
\begin{pf*}{PROOF OF THEOREM~\ref{tensortwoceil}.}
  One way to prove this theorem is to follow the procedures of proving Theorem~\ref{tensorone}. Firstly, we transform the integer matrix completion into an integral maximal flow feasibility problem. However, in the associated $s$-$t$ network, the capacities of arcs between supply nodes and demand nodes are given by the corresponding ceiling charging rates rather than the uniform ones. Next, by similar optimization techniques, we can derive the structure tensor $W(\bm h,\bm r, \bm{\bar{r}},F)$. In the following, we give an alternative proof.

  The necessary part is obvious. When the supply profile is adequate, the supply tails always dominate the demand tails, which implies the nonnegativity of $W(\bm h, \bm r, \bm{\bar{r}},F)$.

  In what follows, we will show the sufficient part. 
  Firstly, apply Lemma~\ref{decomp} to decompose all the terms: $$\left[r_n-\bar{r}_n\left(k_{a_n+1}+k_{a_n+2}+\cdots+k_{d_n}\right)\right]^+,$$ in the formula of~$W(\bm h, \bm r, \bm{\bar{r}},F)$ into
  \begin{align*}
    &p_n\left[q_n+1-\left(k_{a_n+1}+k_{a_n+2}+\cdots+k_{d_n}\right)\right]^+ \\
    +~&(\bar{r}_n-p_n)\left[q_n-\left(k_{a_n+1}+k_{a_n+2}+\cdots+k_{d_n}\right)\right]^+,
  \end{align*}
  where $p_n \in \mathbb{N}$ and $q_n\in \mathbb{N}$ are chosen so that $p_n<\bar{r}_n$ and $r_n=q_n\bar{r}_n+p_n$.

  Then, in view of Theorem~\ref{tensorone}, we observe that there exists a $(0,1)$-matrix $B$ of size~$$\left(\sum_{n=1}^{N}\bar{r}_n\right)\times N,$$ whose column sums are upper bounded by~$\bm h$ and rows can be partitioned into~$N$ groups such that there are $\bar{r}_n$ rows in the $n$th group, for $n=1,2,\ldots,N$, wherein $p_n$ rows have the same row sum~$q_n+1$, $\bar{r}_n-p_n$ rows have the same row sum~$q_n+1$ and all the $\bar{r}_n$ rows are no more than the service time vector corresponding to~$\left(T_{a_n},T_{d_n}\right)$ elementwise. 
  As we can see, the combination of the $\bar{r}_n$ rows can be a power delivery profile satisfying the load requirement $(r_n,\bar{r}_n,a_n,d_n)$. This completes the proof.
\end{pf*}

From the above proof, we can also interpret Theorem~\ref{tensortwoceil} by load decomposition in the sense that we can always decompose a load specified by $(r,\bar{r},a,d)$ into an equivalent collection of sub-loads by Euclidean division as~$r=q\bar{r}+p$. In this collection, there are $p$ sub-loads requiring the same service $(q+1,1,a,d)$, while the remaining $\bar r-p$ sub-loads require the same service $(q,1,a,d)$.
\begin{figure}[t]
\centering
$[0~3~2~3~3]'=\begin{cases}
  [0~1~1~1~1]'\\ \vspace{5pt}
  [0~1~1~1~1]'\\
  [0~1~0~1~1]'
\end{cases}\!\!\!\vline~~
[0~2~2~3~3]'=\begin{cases}
  [0~1~1~1~1]' \vspace{5pt} \\
  [0~0~1~1~1]'\\
  [0~1~0~1~1]'
\end{cases}$
  \caption{Illustration of load decomposition.}
  \label{fig: loaddecomp}
\end{figure}
\begin{exmp}
  Take into account two loads whose ceiling charging rates are both three and service time vectors are both~$[0~1~1~1~1]'$. Their duration requirements are respectively eleven and ten. In Fig.~\ref{fig: loaddecomp}, we respectively give two feasible charging profiles for the loads, denoted by~$[0~3~2~3~3]'$ and~$[0~2~2~3~3]'$. Each original feasible charging profile can be regarded as the combination of three charging profiles where the charging rates are no more than one, as illustrated on the left and right of Fig.~\ref{fig: loaddecomp}. Thus, by decomposing involved loads in the manner described previously, we can reduce rate-constrained differentiated energy services to basic differentiated energy services considered in Section~$4$.
\end{exmp}

Analogously to that discussed in the last section, the minimum purchase profile problem for the rate-constrained differentiated energy services has the following mathematical expression:
\begin{align}
\min_{\bm{p}}\quad& \|\bm p\|_1\tag{Minimum Purchase}\\
\text{s.t.}\quad& \bm p \in \mathbb{N}^T,\nonumber \\
&W(\bm h+\bm p,\bm r, \bm{\bar{r}},F)\geq 0.\label{progapceil}
\end{align}
Likewise, the optimum of the problem, which is defined as the adequacy gap, can be obtained with the help of the associated structure tensor~$W(\bm h, \bm r, \bm{\bar{r}},F)$.
\begin{thm}\label{thmgapceil}
  The optimum of the minimum purchase profile problem~(\ref{progapceil}), i.e., the adequacy gap, is given by \centering $\left |\min\limits_{k_1, k_2,\cdots, k_{\lambda}}W_{k_1k_2\cdots k_{\lambda}}(\bm h,\bm r, \bm{\bar{r}},F)\right|.$
\end{thm}
Algorithm~\ref{alg1} and Algorithm~\ref{alg2} can still be used to obtain a minimum purchase profile after we replace $W(\bm h, \bm r, F)$ with $W(\bm h, \bm r, \bm{\bar{r}},F)$. This is explicit for Algorithm~\ref{alg1}. As for Algorithm~\ref{alg2}, the applicability is verified by Corollary~\ref{coltwo} below, which is an analogy with Lemma~\ref{colone}.
\begin{cor} \label{coltwo}
  The class~$\mathcal{A}(\bm{\hat{h}},\bm r, \bm{\bar{r}},E)$ is nonempty, and~$\bm{\tilde{h}}\prec \bm{\hat{h}}$, then the class~$\mathcal{A}(\bm{\tilde{h}},\bm r, \bm{\bar{r}},E)$ is nonempty.
\end{cor}
\begin{pf}
Since all the involved loads share the same service time, the corresponding pattern matrix~$F$ is~$E$. In this case, the associated tensor $W(\bm{{h}}, \bm r, \bm{\bar{r}},F)$ is reduced to a vector of length~$T+1$: for $k=0,1,\dots,T$,
\begin{equation*}
W_k(\bm h, \bm r, \bm{\bar{r}},F)=\sum_{j>k}^Th_j - \sum_{n=1}^{N}[r_n-\bar{r}_nk]^+.
\end{equation*}
In view of the above formula, we can conclude that if~$W(\bm{\hat{h}}, \bm r, \bm{\bar{r}},F)\geq 0$ and~$\bm{\tilde{h}}\prec \bm{\hat{h}}$, then $W(\bm{\tilde{h}}, \bm r, \bm{\bar{r}},F)\geq 0$ by the definition of the majorization order. The claim then follows by a direct application of Theorem~\ref{tensortwoceil}.
\end{pf}

\section{Conclusions and Future Work}
In this paper, we firstly focus on the matrix completion problem concerning a $(0,1)$-matrix class with given row/column sums and certain zeros prespecified. As a generalization of the classic Gale-Ryser theorem, we use the nonnegativity of an associated structure tensor to characterize the non-emptiness of the considered class. Our simulations demonstrate that the tensor approach can also help find a required matrix more efficiently than existing algorithms, in cases where the pattern matrix has a large number of rows but a simple column structure. Furthermore, we apply the mathematical results to two adequacy problems in the differentiated energy services, namely, the problems of supply/demand matching and the minimum purchase profile. Finally, we consider the more practical rate-constrained differentiated energy services. This extends the results regarding $(0,1)$-matrices to nonnegative integer matrices with prescribed row/column sums, predetermined zeros, and different bounds across the rows.

In the future, more related problems will be studied. From the application perspective, we can take into account the interaction between different loads and the uncertainties in supplies/demands in variants of differentiated energy services. These lead to research on a special~$(-1,0,1)$-matrix class and the nonnegative integer matrix class where row/column sums are not exactly given but described by boundary intervals. In addition, we expect to find more connections between matrix completion problems and partial orders, in particular, the majorization order. Preliminary work in these directions has been presented in~\cite{mo2016duration,mo2017coordinating,mo2018staircase}. 

\begin{ack}                               
The authors thank Prof. Pravin Varaiya of University of California at Berkeley for helpful discussions. 
\end{ack}

\bibliographystyle{ieeetr}        

\begin{thebibliography}{10}

\bibitem{boggs1992resource}
C.~L. Boggs, ``Resource allocation: {E}xploring connections between foraging
  and life history,'' {\em Functional Ecology}, vol.~6, pp.~508--518, 1992.

\bibitem{van1986acquisition}
A.~J. van Noordwijk and G.~de~Jong, ``Acquisition and allocation of resources:
  {T}heir influence on variation in life history tactics,'' {\em The American
  Naturalist}, vol.~128, pp.~137--142, 1986.

\bibitem{ahuja1993network}
R.~K. Ahuja, T.~L. Magnanti, and J.~B. Orlin, {\em Network Flows: {T}heory,
  Algorithms, and Applications}.
\newblock Prentice Hall, 1st~ed., 1993.

\bibitem{krishna2009auction}
V.~Krishna, {\em Auction Theory}.
\newblock Academic Press, 2nd~ed., 2009.

\bibitem{villani2003topics}
C.~Villani, {\em Topics in Optimal Transportation}.
\newblock American Mathematical Society, 2003.

\bibitem{ibaraki1988resource}
T.~Ibaraki and N.~Katoh, {\em Resource Allocation Problems: Algorithmic
  Approaches}.
\newblock MIT Press, 1988.

\bibitem{giua2002firing}
A.~Giua, A.~Piccaluga, and C.~Seatzu, ``Firing rate optimization of cyclic
  timed event graphs by token allocations,'' {\em Automatica}, vol.~38,
  pp.~91--103, 2002.

\bibitem{richert2013optimal}
D.~Richert and J.~Cort{\'e}s, ``Optimal leader allocation in {UAV} formation
  pairs ensuring cooperation,'' {\em Automatica}, vol.~49, pp.~3189--3198,
  2013.

\bibitem{chasparis2016design}
G.~C. Chasparis, M.~Maggio, E.~Bini, and K.-E. {\AA}rz{\'e}n, ``Design and
  implementation of distributed resource management for time-sensitive
  applications,'' {\em Automatica}, vol.~64, pp.~44--53, 2016.

\bibitem{chen2017optimal}
X.~Chen, M.-A. Belabbas, and T.~Ba{\c{s}}ar, ``Optimal capacity allocation for
  sampled networked systems,'' {\em Automatica}, vol.~85, pp.~100--112, 2017.

\bibitem{ghosh1958input}
A.~Ghosh, ``Input-output approach in an allocation system,'' {\em Economica},
  vol.~25, pp.~58--64, 1958.

\bibitem{jain2010efficient}
R.~Jain and J.~Walrand, ``An efficient {N}ash-implementation mechanism for
  network resource allocation,'' {\em Automatica}, vol.~46, pp.~1276--1283,
  2010.

\bibitem{martin2005time}
G.~Mart{\'\i}n-Herr{\'a}n, S.~Taboubi, and G.~Zaccour, ``A time-consistent
  open-loop {S}tackelberg equilibrium of shelf-space allocation,'' {\em
  Automatica}, vol.~41, pp.~971--982, 2005.

\bibitem{gale1957theorem}
D.~Gale, ``A theorem on flows in networks,'' {\em Pacific Journal of
  Mathematics}, vol.~7, pp.~1073--1082, 1957.

\bibitem{ryser1957combinatorial}
H.~J. Ryser, ``Combinatorial properties of matrices of zeros and ones,'' {\em
  Canadian Journal of Mathematics}, vol.~9, pp.~371--377, 1957.

\bibitem{anstee1982properties}
R.~P. Anstee, ``Properties of a class of $(0, 1)$-matrices covering a given
  matrix,'' {\em Canadian Journal of Mathematics}, vol.~34, pp.~438--453, 1982.

\bibitem{anstee1982triangular}
R.~P. Anstee, ``Triangular $(0, 1)$-matrices with prescribed row and column
  sums,'' {\em Discrete Mathematics}, vol.~40, pp.~1--10, 1982.

\bibitem{brualdi2003matrices}
R.~A. Brualdi and G.~Dahl, ``Matrices of zeros and ones with given line sums
  and a zero block,'' {\em Linear Algebra and Its Applications}, vol.~371,
  pp.~191--207, 2003.

\bibitem{chen2016constrained}
W.~Chen, Y.~Mo, L.~Qiu, and P.~Varaiya, ``Constrained $(0, 1)$-matrix
  completion with a staircase of fixed zeros,'' {\em Linear Algebra and Its
  Applications}, vol.~510, pp.~171--185, 2016.

\bibitem{brualdi2006combinatorial}
R.~A. Brualdi, {\em Combinatorial Matrix Classes}.
\newblock Cambridge University Press, 1st~ed., 2006.

\bibitem{olkin2016inequalities}
A.~W. Marshall, I.~Olkin, and B.~C. Arnold, {\em Inequalities: {T}heory of
  Majorization and Its Applications}.
\newblock Springer, 2nd~ed., 2011.

\bibitem{berger2011dag}
A.~Berger and M.~M{\"u}ller-Hannemann, ``Dag realizations of directed degree
  sequences,'' in {\em International Symposium on Fundamentals of Computation
  Theory}, pp.~264--275, Springer, 2011.

\bibitem{chen1966realization}
W.-K. Chen, ``On the realization of a $(p, s)$-digraph with prescribed
  degrees,'' {\em Journal of the Franklin Institute}, vol.~281, pp.~406--422,
  1966.

\bibitem{erdosgallai1960degree}
P.~Erd{\"o}s and T.~Gallai, ``Graphs with prescribed degrees of
  vertices~({H}ungarian),'' {\em Matematikai Lapok}, pp.~264--274, 1960.

\bibitem{siano2014demand}
P.~Siano, ``Demand response and smart grids~--~{A} survey,'' {\em Renewable and
  Sustainable Energy Reviews}, vol.~30, pp.~461--478, 2014.

\bibitem{mo2017differentiated}
Y.~Mo, W.~Chen, and L.~Qiu, ``Differentiated energy services: {M}ultiple
  arrival times and multiple deadlines,'' in {\em 20th World Congress of the
  International Federation of Automatic Control (IFAC)}, pp.~207--212, 2017.

\bibitem{korte2012combinatorial}
B.~Korte and J.~Vygen, {\em Combinatorial Optimization: Theory and Algorithms}.
\newblock Springer, 6th~ed., 2018.

\bibitem{trevisan2011combinatorial}
L.~Trevisan, {\em Combinatorial Optimization: {E}xact and Approximate
  Algorithms}.
\newblock Standford University, 2011.

\bibitem{dantzig1955max}
G.~B. Dantzig and D.~R. Fulkerson, ``On the max-flow min-cut theorem of
  networks,'' Tech. Rep. {P}-826, RAND Corporation, 1955.

\bibitem{ford1956maximal}
L.~R. Ford~Jr. and D.~R. Fulkerson, ``Maximal flow through a network,'' {\em
  Canadian Journal of Mathematics}, vol.~8, pp.~399--404, 1956.

\bibitem{fulkerson1959network}
D.~R. Fulkerson, ``A network-flow feasibility theorem and combinatorial
  applications,'' {\em Canadian Journal of Mathematics}, vol.~11, pp.~440--451,
  1959.

\bibitem{mirsky1968combinatorial}
L.~Mirsky, ``Combinatorial theorems and integral matrices,'' {\em Journal of
  Combinatorial Theory}, vol.~5, pp.~30--44, 1968.

\bibitem{edmonds1972theoretical}
J.~Edmonds and R.~M. Karp, ``Theoretical improvements in algorithmic efficiency
  for network flow problems,'' {\em Journal of the ACM (JACM)}, vol.~19,
  pp.~248--264, 1972.

\bibitem{goldberg1988new}
A.~V. Goldberg and R.~E. Tarjan, ``A new approach to the maximum-flow
  problem,'' {\em Journal of the ACM (JACM)}, vol.~35, pp.~921--940, 1988.

\bibitem{orlin1997polynomial}
J.~B. Orlin, ``A polynomial time primal network simplex algorithm for minimum
  cost flows,'' {\em Mathematical Programming}, vol.~78, pp.~109--129, 1997.

\bibitem{boykov2004experimental}
Y.~Boykov and V.~Kolmogorov, ``An experimental comparison of min-cut/max-flow
  algorithms for energy minimization in vision,'' {\em {IEEE} Transactions on
  Pattern Analysis and Machine Intelligence}, vol.~26, pp.~1124--1137, 2004.

\bibitem{nayyar2016duration}
A.~Nayyar, M.~Negrete-Pincetic, K.~Poolla, and P.~Varaiya,
  ``Duration-differentiated energy services with a continuum of loads,'' {\em
  {IEEE} Transactions on Control of Network Systems}, vol.~3, pp.~182--191,
  2016.

\bibitem{mo2017coordinating}
Y.~Mo, W.~Chen, and L.~Qiu, ``Coordinating flexible loads via optimization in
  the majorization order,'' in {\em 56th {IEEE} Conference on Decision and
  Control (CDC)}, pp.~3495--3500, 2017.

\bibitem{negrete2016rate}
M.~Negrete-Pincetic, A.~Nayyar, K.~Poolla, F.~Salah, and P.~Varaiya,
  ``Rate-constrained energy services in electricity,'' {\em {IEEE} Transactions
  on Smart Grid}, vol.~9, pp.~2894--2907, 2018.

\bibitem{mo2016duration}
Y.~Mo, W.~Chen, and L.~Qiu, ``Duration-differentiated energy services with
  peer-to-peer charging,'' in {\em 55th {IEEE} Conference on Decision and
  Control (CDC)}, pp.~7514--7519, 2016.

\bibitem{mo2018staircase}
Y.~Mo, W.~Chen, and L.~Qiu, ``Staircase pattern constrained zero-one matrix
  completion with uncertainties and its applications,'' in {\em 23rd
  International Symposium on Mathematical Theory of Networks and Systems
  (MTNS)}, pp.~222--225, 2018.

\end{thebibliography}




\appendix

\end{document}